\documentclass[final,5p,times,twocolumn]{elsarticle}

\usepackage{amssymb}
\usepackage{amsmath}
\usepackage{listings}
\usepackage{hyperref}
\usepackage{cancel,soul,ulem}
\usepackage{listings}
\usepackage{xcolor}
\usepackage{tcolorbox}
\usepackage{array}      
\usepackage{hyperref}   
\usepackage{graphicx}   
\usepackage{booktabs}   
\usepackage{caption}    
\usepackage{multirow}  
\usepackage{makecell}  
\usepackage{amssymb}

\tcbuselibrary{breakable}

\newtcolorbox{insight}
{colback=gray!20,colframe=gray,boxrule=0.5pt,boxsep=0mm,leftupper=1mm,rightupper=1mm}
\newtcolorbox{prompt}{colback=yellow!20,colframe=black,boxrule=1pt,boxsep=0mm,leftupper=1mm,rightupper=1mm,breakable}
\newtcolorbox{respond}{colback=green!20,colframe=black,boxrule=1pt,boxsep=0mm,leftupper=1mm,rightupper=1mm,breakable}
\usepackage{pifont}
\newcommand{\added}[1]{\textcolor{black}{#1}}

\journal{ The Journal of Systems \& Software}

\begin{document}

\begin{frontmatter}



\title{An Empirical Study on Low-Code Programming using Traditional vs Large Language Model Support}


\author[SYSUSE]{Yongkun Liu}
\ead{liuyk39@mail2.sysu.edu.cn}

\author[SYSUSE]{Jiachi Chen\corref{cor1}}
\ead{chenjch86@mail.sysu.edu.cn}

\author[Melbourne,WA]{Tingting Bi}
\ead{Tingting.Bi@unimelb.edu.au}

\author[Monash]{John Grundy}
\ead{John.Grundy@monash.edu}

\author[SYSUSE]{Yanlin Wang}
\ead{wangylin36@mail.sysu.edu.cn}

\author[SYSUAI]{Jianxing Yu}
\ead{yujx26@mail2.sysu.edu.cn}

\author[CSE]{Ting Chen}
\ead{brokendragon@uestc.edu.cn}

\author[Glasgow]{Yutian Tang}
\ead{yutian.tang@glasgow.ac.uk}

\author[SYSUSE,SYSUSE2]{Zibin Zheng}
\ead{zhzibin@mail.sysu.edu.cn}

\cortext[cor1]{Corresponding author}

\affiliation[SYSUSE]{organization={School of Software Engineering, Sun Yat-sen University},
            country={China}}

\affiliation[Melbourne]{organization={The University of Melbourne},
            country={Australia}}

\affiliation[WA]{organization={The University of Western Australia},
            country={Australia}}

\affiliation[Monash]{organization={Faculty of Information Technology, Monash University},
            country={Australia}}

\affiliation[SYSUAI]{organization={School of Artificial Intelligence, Sun Yat-sen University},
            country={China}}

\affiliation[CSE]{organization={Shenzhen Institute for Advanced Study, University of Electronic Science and Technology of China},
            country={China}}

\affiliation[Glasgow]{organization={University of Glasgow},
            country={United Kingdom}}

\affiliation[SYSUSE2]{organization={Zhuhai Key Laboratory of Trusted Large Language Models},
            country={China}}
            


\begin{abstract}
Low-code programming (LCP) refers to programming using models at higher levels of abstraction, resulting in less manual and more efficient programming, and reduced learning effort for amateur developers.
Many LCP tools have rapidly evolved and have benefited from the concepts of visual programming languages (VPLs) and programming by demonstration (PBD). With the huge increase in interest in using large language models (LLMs) in software engineering, LLM-based LCP has begun to become increasingly important.
However, the technical principles and application scenarios of traditional approaches to LCP and LLM-based LCP are significantly different. Understanding these key differences and characteristics in the application of the two approaches to LCP by users is crucial for LCP providers in improving existing and developing new LCP tools and in better assisting users in choosing the appropriate LCP technology.
We conducted an empirical study of both traditional LCP and LLM-based LCP. We analyzed developers' discussions on Stack Overflow (SO) over the past three years and then explored the similarities and differences between traditional LCP and LLM-based LCP features and developer feedback.
Our findings reveal that while traditional LCP and LLM-based LCP share common primary usage scenarios, they significantly differ in scope, limitations, and usage throughout the software development lifecycle, particularly during the implementation phase. We also examine how LLMs impact and integrate with LCP, discussing the latest technological developments in LLM-based LCP, such as its integration with VPLs and the application of LLM Agents in software engineering.
\end{abstract}



\begin{keyword}


Low-code programming \sep Large language model \sep Empirical Study

\end{keyword}

\end{frontmatter}



\section{Introduction}
\label{Introduction}

Low-code programming (LCP) has recently become a discussed topic in software development communities~\cite{bock2021low}. It primarily aims to minimize the use of low-level textual programming languages and utilize higher levels of model abstractions that align more closely with natural human thought processes for developing software~\cite{hirzel2023low}.
By applying LCP, professional developers can reduce time-consuming and repetitive manual coding, and thereby improve development efficiency. LCP also allows amateur developers, referring to individuals with minimal or no formal programming training, to achieve programming objectives with much lower learning costs and effort \cite{sahay2020supporting,bragancca2021towards}.

Recently, Hirzel \cite{hirzel2023low} categorized the latest LCP technologies into three primary technologies: visual programming languages (VPLs) \cite{burnett1995visual}, programming by demonstration (PBD) \cite{cypher1993watch}, and programming by natural language (PBNL) \cite{jiang2022discovering}.
VPLs allows the user to create programs by manipulating program elements graphically. 
PBD means that the system can record tasks performed manually by the user via a keyboard and mouse, subsequently generating programs that perform replicate these actions.
PBNL translates user-provided natural language text into an executable program.
The main technologies currently applied by popular LCP tools in the industry were VPLs and PBD. 
For example, Microsoft PowerApps \cite{powerapps} uses VPLs to enable users to design user interfaces and workflows using drag-and-drop visualization modules, bypassing the need for manual programming.
Uipath \cite{uipath} employs PBD to automate processes. With its built-in logger, users can capture UI mouse movements and keyboard activities, generating precise automation scripts.

The recent emergence of large language models (LLMs) has validated the capability of PBNL-based approaches to assist software development 
\cite{brown2020language}. LLMs are widely used to generate code across various programming languages using natural language and achieve great performance on various programming tasks, e.g., code generation and code completion~\cite{poldrack2023ai,bubeck2023sparks}, which significantly enhances  software development efficiency. We define this type of LCP as LLM-based LCP. 

With the growing popularity of LLMs, PBNL has also gained significant attention. Developers can now engage with LLMs like ChatGPT to implement LCP by using natural language. In this paper, we refer to LCP involving prevalent VPLs and PBD as ``\textbf{\textit{traditional LCP}}''. In contrast, LCP based on LLMs and programming by natural language is termed as ``\textbf{\textit{LLM-based LCP}}''.
\added{To enhance clarity and ensure consistent terminology, we include a glossary (Table \ref{tab:terminology_glossary}) that defines and summarizes the key concepts used throughout this paper—namely Traditional LCP, LLM-based LCP, VPL, PBD, and PBNL—and explicitly notes their areas of overlap.}

\begin{table}[t]
  \centering
  \caption{Terminology and Definitions}
  \resizebox{0.5\textwidth}{!}{
  {\color{black}
    \begin{tabular}{m{3.5cm}m{7cm}}
    \toprule
    \textbf{Term} & \textbf{Definition} \\
    \midrule
    Traditional LCP & Platforms that reduce coding effort through components, visual interfaces, or demonstration, covering PBNL, VPL, and PbD as subcategories. \\
    \midrule
    LLM-based LCP & Low-code platforms powered by large language models, often combined with PBNL or VPL to support natural language or visual programming. \\
    \midrule
    PBNL& Programming by Natural Language: creating programs via natural language instructions; a subcategory of Traditional LCP and a core mechanism in many LLM-based LCPs. \\
    \midrule
    VPL& Visual Programming Language: programming through graphical interfaces (e.g., drag-and-drop, flowcharts); a common form of Traditional LCP that can also integrate with LLMs. \\
    \midrule
    PbD & Programming by Demonstration: creating programs by user demonstrations that the system abstracts into logic; a subcategory of Traditional LCP. \\

    \bottomrule
    \end{tabular}%
    }
  }
  \label{tab:terminology_glossary}%
\end{table}%


Both ``traditional LCP'' and ``LLM-based LCP'' have their advantages and disadvantages. 
Traditional LCP is intuitive, readable, and unambiguous for users, which significantly reduces syntax errors and some simple programming errors. However, it usually has limited application scenarios. For example, PowerApps can only be used for web development.
In contrast, LLM-based LCP, with its vast repository of programming knowledge encompassing scripts, algorithms, and more, offers a wide range of applications. 
This makes it more accessible for amateur developers to create applications. Nevertheless, the generated programs may suffer from inaccuracies due to mis-understanding user prompts, the hallucination problem of LLMs, and erroneous generated code, leading to errors.

To understand the key characteristics and limitations of traditional LCP and LLM-based LCP, we analyze the posts on Stack Overflow (SO) \cite{StackOverflow}, which is the most popular and widely used question and answer (Q\&A) platform for developers to ask and answer questions.
We first applied 27 keywords related to traditional LCP and 8 keywords pertinent to LLM-based LCP to preliminary filter out related posts. This process produced a substantial data set of 7,367 posts on traditional LCP and 7,468 posts on LLM-based LCP \added{over} the last three years. To further refine our dataset and ensure relevance to our research objectives, we employed ChatGPT, utilizing custom-designed prompts. Finally, we performed a manual analysis on the 1,688 posts filtered (642 and 1,046 posts for traditional and LLM-based LCP, respectively) \cite{glaser1965constant} to answer the following three research questions:

\noindent \textbf{RQ1.What application domains do discussions on traditional LCP and LLM-based LCP focus on?}

Answering this RQ enhances our understanding of the distinct characteristics of the two LCPs and provides insights into their primarily usage by current users. The analysis for this RQ was conducted using data from 470 posts related to traditional LCP and 566 posts pertaining to LLM-based LCP. Our findings reveal that LLM-based LCP covers a broader spectrum of application domains \added{compared} to traditional LCP, many of which focus on web development. Users typically employ LLM-based LCP for addressing general programming challenges, whereas traditional LCP is \added{used more frequently to resolve} API integration issues.

\noindent \textbf{RQ2. To which software development tasks do traditional LCP and LLM-based LCP contribute?}

There are several phases in \added{the} software development life cycle, i.e., requirement analysis \& planning, design, implementation, deployment, testing, \added{maintenance}.
It is essential to determine the stages in which developers are more actively involved in discussions about traditional LCP and LLM-based LCP. Such insights can reveal where each LCP type offers the most benefit.
We used \added{the} constant comparison method to collect  412 instances related to traditional LCP, and 559 instances related to LLM-based LCP. Subsequently, we classified them into their respective software development life cycle phases. We found that both traditional LCP  and LLM-based LCP discussions focused on the implementation phase, and the former focused more on the deployment phase than the latter.

\noindent \textbf{RQ3. What are the limitations associated with traditional LCP and LLM-based LCP?}

Understanding the limitations encountered by users of \added{the} Traditional LCP and LLM-based LCP respectively is critical. This knowledge is instrumental for LCP providers in shaping the development of future tools. In addressing this research question, we analyzed a total of 65 instances related to LLM-based LCP and 95 instances concerning traditional LCP. From this analysis, we identified seven limitations for traditional LCP and five for LLM-based LCP, as reported in SO data.
Our comparative analysis identified the most critical limitations suggested for each LCP type. For instance, users of LLM-based LCP grapple with higher demands for technical expertise and face more pronounced reliability concerns. Such insights are invaluable for understanding the distinct challenges posed by each LCP type, guiding providers in tailoring their development strategies to better meet user needs and expectations.

In summary, our main contributions are as follows:
\vspace{-0.2cm}
\begin{itemize} 
	\item We compare LLM-based LCP with traditional LCP and identified their unique characteristics, application domains and limitations. 

        
        \item We discuss how LLMs drive the development of traditional LCP tools, how the concept of VPLs drives the development of LLM-based LCP, and the impact of LLMs agent on LCP. 

	\item We publish our dataset and results at \href{https://zenodo.org/records/11232842}{https://zenodo.org/records/11232842} to facilitate further studies.
\end{itemize}

The rest of this paper is structured as follows.
We introduce some background knowledge and concepts in Section \ref{sec:background}.
In Section \ref{Methodology}, we elucidate the methodologies employed for investigating LCP. Section \ref{Results} unveils the results and noteworthy findings gleaned from our thorough exploration into LCP. Section \ref{Discussions} provides an in-depth discussion of the latest LLM-based LCP technologies and future developments. Section \ref{related} introduces related work. The conclusion and future \added{work} are articulated in Section \ref{conclusion}.

\section{Background}
\label{sec:background}

\subsection{Low-Code Programming}
The term ``low-code'' first appeared in a report published by Forrester Research in 2014 \cite{F2014}.
The original meaning of the term was that applications could be developed quickly with minimal manual coding and minimal up-front investment in setup, training, and deployment.
However, in subsequent reports and research on LCP, the definition of the term ``low-code'' remained in a constant state of flux and ambiguity \cite{richardson2016forrester,rymer2017vendor,di2022low}.
\added{To address this ambiguity, the definition proposed by Hirzel~\cite{hirzel2023low} is adopted in this paper, which emphasizes that LCP are technologies designed to minimize text-based programming, with the core aim of reducing reliance on manual coding, whether through VPLs, PBD, or PBNL.}

Whether for professional developers or amateur developers, the purpose of LCP is to save manual programming time and reduce the learning cost of manual programming as much as possible \cite{van2018robotic}.
Traditional LCP tools, such as Power Apps \cite{powerapps} and \added{OutSystems} \cite{outsystems}, typically include a visual development interface for drag-and-drop development. \added{They are often equipped with cloud services for automated deployment, but cloud deployment is a common feature rather than a defining requirement of LCP.}

\subsection{Large Language Models}
LLMs, such as CoPilot \cite{copilot} and ChatGPT \cite{chatgpt}, refer to deep learning decoder models trained on massive amounts of text, and increasingly other data like images and models, typically based on the Transformer architecture~\cite{vaswani2017attention}. 
In the realm of software development, LLMs have emerged as transformative tools, showcasing remarkable capabilities across various tasks, including code repair, code generation, test case generation, documentation and code summarization \cite{hou2023large,ma2023scope}.
A key example of an AI-assisted programming tool based on LLMs is \added{GitHub} Copilot \cite{copilot}.
By interpreting comments and understanding the context of the code being written, \added{GitHub} Copilot is capable of suggesting relevant code snippets and even completing entire code, thereby expediting the coding process and reducing the learning cost on developers.
A number of \added{LLM} agents have also been proposed specifically for software generation recently, which can generate code and software throughout the software development life cycle \cite{qian2023communicative,hong2023metagpt}. This shows that LLMs show strong potential in the LCP field.

\subsection{Prompt Design}
In this paper, ``prompt design'' refers to crafted text descriptions used to guide LLMs (e.g., ChatGPT) in generating their output. Recent LLMs can engage in  multi-turn conversations out of the box and accomplish diverse, complex tasks specified in the input text. However, research has shown that the quality of LLMs' responses and user satisfaction can  be strongly influenced by prompt design \cite{zamfirescu2023johnny}. The same communicative intent may elicit comprehensive, detailed, and helpful responses or responses that are unhelpful or even incorrect, depending entirely on how the prompt is designed. Therefore, designing a well-crafted prompt is crucial for effectively guiding LLMs in completing a given task.

\subsection{Software Development Processes}
The software development life cycle (SDLC) is used to plan and manage the process of software development \cite{shylesh2017study}. It typically involves breaking down the software development work into smaller, parallel, or sequential steps or subprocesses to improve the development process. There are many models of SDLC, usually encompassing  key tasks including planning, requirements engineering, design, implementation, testing, deployment, and \added{maintainance} tasks.
Software project planning includes team formation, tool selection, estimation, contracting and others.
Requirements engineering involves \added{understanding} and document the client's needs and planning the overall strategy for the software development process.
Requirements engineering tasks involve understanding and documenting the client's needs, as well as planning the overall strategy of the software development process.
Design tasks include developers deciding the architecture and design of the software, including interface, \added{behaviour} and data, providing detailed explanations of how it will meet the requirements.
During implementation tasks, developers code the software product.
Testing tasks include the process of verifying and confirming the software to ensure it meets the requirements and has no vulnerabilities.
Deployment involves the release of the software and making it available, typically involving installation and configuration processes.
Finally, there are maintenance tasks, where developers may update the software, fix any issues that arise, and possibly add new features or functionalities.

\section{Methodology}
\label{Methodology}
\begin{figure*}[t]
  \centering
  \includegraphics[width = 1.0\textwidth]{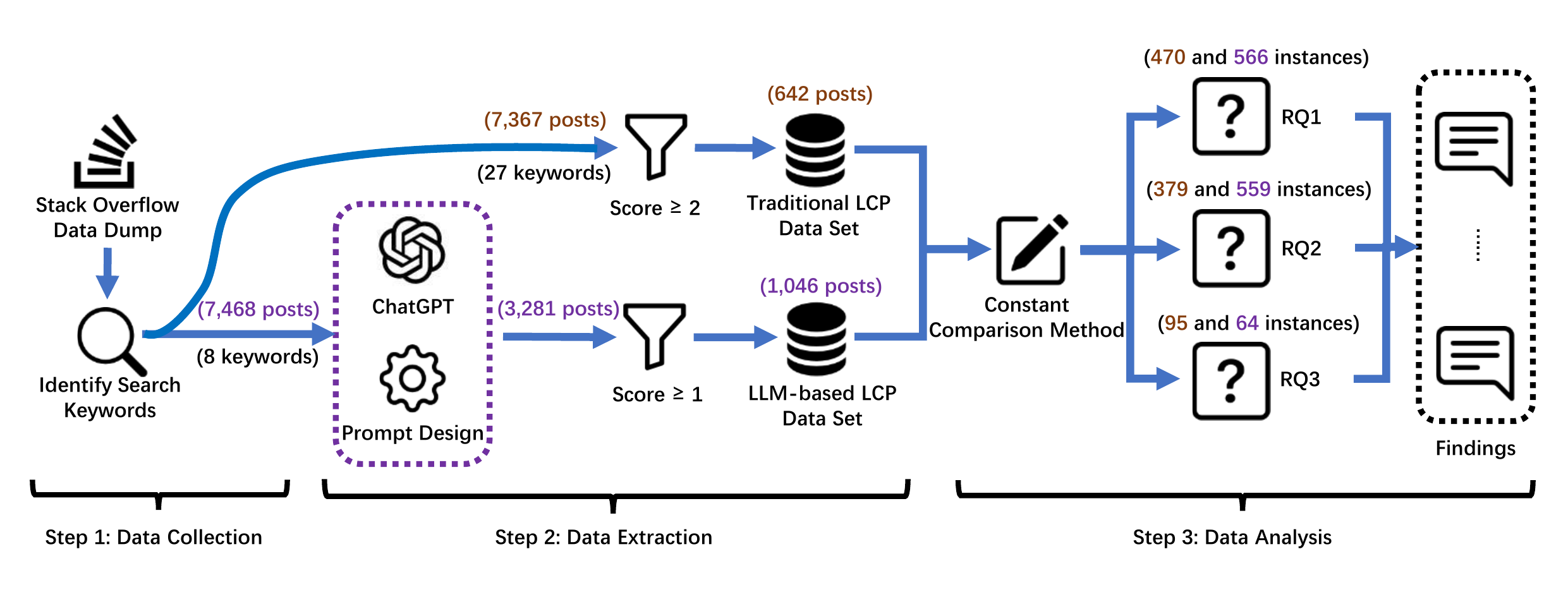}
  \captionsetup{skip=-5pt}
  \caption{Overview of the investigation process, each step corresponds to a subsection in Section \ref{Methodology}}
  \label{fig:Overview}
\end{figure*}

Figure \ref{fig:Overview} shows an overview of the process used to derive answers to our three research questions (RQs) from Stack Overflow (SO) posts. 
Step 1 involves downloading the entire historical dataset of posts from Stack Overflow. We then use a set of specifically chosen keywords to \added{preliminarily} filter LCP-related posts.
In Step 2 we focus on the design of effective prompts and apply ChatGPT to further refine the data. This step ensures the removal of irrelevant data, ultimately obtaining the traditional LCP dataset with 642 posts and the LLM-based LCP dataset with 1,046 posts.
In Step 3 we use the constant comparison method to manually analyze the data and obtain the answer of the three RQs.

\vspace{-3 mm}
\subsection{Step 1: Data Collection}
\label{Data_Collection}

\subsubsection{SO Posts Collection}

We downloaded the entire SO data dump \cite{SODump} of September 2023, which includes content contributed by users on the SO network. The SO data dump consists of five XML documents: Badges.xml, Comments.xml, Users.xml, Posts.xml, and Votes.xml.
For our purposes, we used the Posts.xml data file, which contains comprehensive details of all posts, including the unique ID of each post, type (question or answer), title, etc.
The data dump included posts spanning from July 2008 to September 2023.
Notably, the groundbreaking GPT-3~\cite{brown2020language}  LLM was released in June 2020, and prior to this, there was little discussion on LLM on SO. To ensure the timeliness and consistency of the data supporting the research results, we utilized data from the last three years (June 2020 to September 2023).

\subsubsection{Keywords-Based Post Filtering.}

After collecting all historical posts, we conducted keyword-based filtering to select the data.
For traditional LCP, we used the \added{term} ``low-code'' and a list of terms relevant to some popular LCP tools as keywords. For LLM-based LCP, we employed eight LLM-related keywords for further filtering. 
Additionally, we applied case-ambiguity matching in our keyword filtering to enhance the inclusivity of our search.

\begin{itemize}
    \item \textbf{Keywords for traditional LCP}: To capture the discussions about traditional LCP by developers, we searched not only for the term ``low-code'' but also included names of top low-code tool. This approach is based on the observation that developers often mention specific tool names when discussing difficulties encountered while using LCP tools on SO. We compiled a list of top low-code tools from a previous work~\cite{al2021empirical} and the top 10 platforms \added{from G2} website (A popular website to review software) \cite{g2}. The list of top low-code tools contained 24 tool names. \added{We deliberately excluded the term ``no code'' as a keyword, since it often introduces substantial noise (e.g., ``no code available'', ``no code error'') and yields very few relevant posts, as also reported by Luo et al.~\cite{luo2021characteristics}.} The final set of keywords \added{was} as follows: 

\begin{lstlisting} [escapeinside={(*}{*)}, caption={Keywords related to traditional LCP}, label = {code:Keywords_traditional}]
["Zoho Creator", "App Maker", "Salesforce Lightning", "Quickbase", "OutSystems", "Mendix", "Vinyl", "Appian", "PowerApps", "UiPath", "Claris Filemaker", "Servicenow", "Salesforce",  "IAR Embedded Workbench", "Airtable", "Pega", "Appy Pie", "Glide", "Nintex", "AppSheet", "AppMySite", "Softr", "Ninox", "Quixy", "Low Code", "Low-Code", "Lowcode"]
\end{lstlisting}

    \item \textbf{Keywords for LLM-based LCP}: To identify posts related to LLM-based low-code programming, we used ``LLM'' and a list of top LLM-based tools as keywords. Note that many posts retrieved \added{in} this way were not directly relevant to LCP, thus we \added{needed} further filtering, which \added{is} detailed in Section \ref{Extract_GPT}. Based on the latest works \cite{hadi2023survey,teubner2023welcome}, we selected the most popular LLM-based tools and code generation tools as keywords. The final search keywords included:

\begin{lstlisting} [escapeinside={(*}{*)}, caption={Keywords related to LLM-based LCP}, label = {code:Keywords_LLM}]
["LLM", "LLMs", "GPT", "ChatGPT", "Copilot", "Bard", "Codex", "Claude"]
\end{lstlisting}
\end{itemize}
This keyword-based approach to analyzing SO data from the past three years yielded a substantial number of posts: 7,367 related to traditional LCP and 7,468 pertaining to LLM.

\subsection{Step 2: Data Extraction using ChatGPT.}
\label{Extract_GPT}

After the above steps, we obtained the preliminary dataset related to traditional LCP and LLM-based LCP. 
However, the dataset may contain many posts irrelevant to LCP. For example, some posts \added{that} contain LLM-related keywords may discuss the fine-tuning process of LLMs and not usage for LCP.
Given the large volume of such data, manual data filtering can be both time-consuming and error-prone. Since ChatGPT \cite{chatgpt} has demonstrated excellent natural language comprehension and text classification capabilities \cite{orru2023human}, we \added{employed} ChatGPT to help filter data.
\subsubsection{Prompt Design}

We optimized our ChatGPT prompt strategy based on some existing prompt strategies \cite{park2023generative,xu2023expertprompting,wei2022chain}.
Specifically, we first let ChatGPT analyze each post data set sentence to ascertain its relevance to LCP. Then, ChatGPT was instructed to deliver an answer (True or False) based on the analysis results, which facilitates large-scale data filtering.
The prompt design details are shown below:

\vspace{-2mm}
\begin{prompt}
\textit{\textbf{User Prompt 1:}} 
You are \textbf{[ROLE]}. \textbf{[KNOWLEDGE DESCRIPTION]}. \textbf{[TASK DESCRIPTION]}. Think step by step, carefully. The input statements are as follow: \textbf{[INPUT]}.
\hspace{1em}
\end{prompt}

\vspace{-1mm}

\vspace{-1mm}
\begin{respond}
\textit{\textbf{ChatGPT Response 1:} }
\textbf{[ANALYSIS]}.
\end{respond}
\vspace{-1mm}

\vspace{-1mm}
\begin{prompt}
\textit{\textbf{User Prompt 2:} }
\textbf{[OUTPUT FORMAT]}.
\end{prompt}
\vspace{-1mm}

\vspace{-1mm}
\begin{respond}
\textit{\textbf{ChatGPT Response 2:} }
\textbf{[TRUE/FALSE]}.
\end{respond}
\vspace{-1mm}

where the placeholder \textbf{[ROLE]} in \textbf{\textit{User Prompt 1}} denotes the specific role assigned to ChatGPT. In this study, we designated the role as an expert in the field of software engineering. In addition, we described the concepts and knowledge about LCP in the placeholder \textbf{[KNOWLEDGE DESCRIPTION]}, ensuring that ChatGPT has accurate prior knowledge to accomplish the task. We assigned a task to ChatGPT in the placeholder \textbf{[TASK DESCRIPTION]}, which involved analyzing sentence by sentence whether the input content \added{was} related to LCP. Finally,\added{we} entered the post to be assessed in the placeholder \textbf{[INPUT]}.
ChatGPT then returned the result of a sentence-by-sentence analysis \textbf{[ANALYSIS]} in \textbf{\textit{ChatGPT Response 1}}.
In the placeholder \textbf{[OUTPUT FORMAT]} of \textbf{\textit{User Prompt 2}}, we required ChatGPT to assess whether each sentence in the \textbf{[ANALYSIS]} is related to LCP. Output \textbf{TRUE} if any sentence is relevant, otherwise output \textbf{FALSE}.
The full text of \added{the} prompt design and results \added{is} available in our open source repository.

\subsubsection{Filter by Posts' Score}

Through our data filtering using ChatGPT, we reduced the dataset of LLM-based LCP from 7,468 posts to 3,281 posts.
However, this is still a large volume for human analysis. To facilitate higher quality data selection, we ranked posts based on their scores in SO~\footnote{\added{The score of a post is determined by user votes in SO and is calculated as $Score = Upvotes - Downvotes$, which reflects both content quality and community recognition.}}. 
A higher score typically indicates a higher quality discussion.
In this paper, we applied score thresholds of score$\geq$1 and score$\geq$2 for the traditional LCP dataset and LLM-based LCP dataset, respectively. 
\added{The rationale is twofold: (1) posts with non-positive scores were excluded to remove low-quality content; and (2) to ensure fairness, we ranked posts by score and included not only the top 500 but also all posts tied at the cutoff score.}
After this additional filtration, our datasets comprised 642 traditional LCP posts and 1,046 LLM-based LCP posts.

\subsection{Step 3: Data Analysis}
\added{To analyze the collected data qualitatively, we employed the constant comparison method~\cite{glaser1965constant}. This method was originally proposed by Glaser and Strauss and has been widely applied in qualitative research. Its core idea is to continuously compare new and existing data during the analysis process, thereby gradually deriving higher-level concepts and categories.
In this study, our application of the constant comparison method consisted of the following steps:}
(1) In the initial phase, the first author summarized the collected data, aligning it with the specific RQs. This involved assigning codes to individual posts, capturing the content's essence. 
(2) Subsequently, the first author aggregated these codes into higher-level concepts. These concepts were then transformed into overarching categories, fostering a more abstract representation of the information. 
(3) To enhance the reliability and robustness of the analysis, the second author meticulously reviewed the assigned codes and the derived categorizations. \added{We calculated Cohen’s Kappa coefficient to quantify inter-rater agreement, which reached 0.92, indicating strong consistency between coders. In cases of disagreement, the two authors discussed until consensus was reached; if consensus could not be immediately achieved, the disagreement was revisited until alignment was obtained.}
(4) To effectively code and categorize data, we used the qualitative data analysis tool MAXQDA \cite{maxqda}, a tool that facilitates coding text and abstracting these codes into conceptual categorizations.

It is worth noting that despite the use of keywords and ChatGPT for filtering, some posts unrelated to LCP may still be present. Fortunately, all filtered data \added{underwent} a manual analysis process, during which we \added{removed} any irrelevant posts. For instance, in post \textit{\href{https://stackoverflow.com/questions/76965127}{A76965127}}, the user discussed the robustness of LLM's embedding vector representations to subtle differences, which was unrelated to LLM-based LCP.
After manual review and labeling of two datasets, we identified 451 and 230 posts unrelated to LCP, respectively. Consequently, in the traditional LCP  and LLM-based LCP datasets, the actual number of posts discussing LCP is 411 and 595, respectively.
\added{This means that at the ChatGPT screening stage, the accuracy of ChatGPT in identifying LCP-related posts was 56.9\%. While the precision was relatively low due to our intentional prompt design favoring higher recall, the method achieved a recall of 87.18\%, which ensured that most potentially relevant posts were captured. Although this introduced some false positives, the subsequent manual verification effectively corrected them, thereby mitigating the impact of noise. As a result, the final datasets reached both a reliable quality and a sufficient scale to support further qualitative analysis.}

\section{Results}
\label{Results}
\label{sec:Results}


\begin{table*}[t]
  \centering
  \caption{Distribution of the SO discussion on traditional LCP across application areas}
  \resizebox{1.0\textwidth}{!}{
    \begin{tabular}{m{3cm}<{\centering}|m{2.2cm}<{\centering}|m{6cm}|m{8cm}}
    \hline
    \textbf{Application Areas} & \multicolumn{1}{p{4.04em}|}{\textbf{Proportion(\%)}} & \qquad \quad \textbf{Description} & \qquad \qquad \qquad \qquad \qquad \textbf{Example} \\
    \hline
    API integration & 36.60  &  Discussion related to using API functionality within the LCP tools. & \textit{\href{https://stackoverflow.com/questions/67883356}{Q67883356}: I use the client ID (=AppID) and token in the ArcGIS JavaScript API like below:...Only implementing this gives me an error...} \\
    \hline
    Web Backend & 24.26  &   Discussion related to building logical flows, function usage, and authentication within LCP tools. & \textit{\href{https://stackoverflow.com/questions/63850758}{Q63850758}: I'm currently attempting to build an Airtable-esque filter component.} \\
    \hline
    Data Management & 23.40  &   Discussions related to database management and data manipulation. & \textit{\href{https://stackoverflow.com/questions/63319755}{Q63319755}: How to use LINQ on a DataTable in Uipath?...}  \\
    \hline
    Web Customization & 15.74  &   Discussion related to UI, buttons, and component customization. & \textit{\href{https://stackoverflow.com/questions/62646468}{Q62646468}: I am able to trigger the first button click and I am able to cause an alert with the second button click, but for some reason when I try to trigger the first button click in the click event handler of the second button, it doesn't work.} \\
    \hline
    \end{tabular}%
    }
  \label{tab:RQ1-Traditional}%
\end{table*}%

\begin{table*}[t]
  \centering
  \caption{Distribution of the SO discussion on LLM-based LCP across application areas}
  \resizebox{1.0\textwidth}{!}{
    \begin{tabular}{m{3cm}<{\centering}|m{2.2cm}<{\centering}|m{6cm}|m{8cm}}
    \hline
    \textbf{Application Areas} & \multicolumn{1}{p{4.04em}|}{\textbf{Proportion(\%)}} & \qquad\quad \textbf{Description} & \qquad \qquad \qquad \qquad \qquad \textbf{Example} \\
    \hline
    General & 32.86  & General issues encountered during development using LCP, such as function format, regular expression, debugging and so on.
 & \textit{\href{https://stackoverflow.com/questions/76930839}{A76930839}: When asked correctly at ChatGPT it answers:I can see the issue in your code. There is a small typo in the RETURN statement for the Celsius conversion. You have misspelled the variable name temperature as temperature.}  \\
    \hline
    Web Frontend & 16.78  & Discussion of elements related to HTML, CSS, UI interfaces, etc. & \textit{\href{https://stackoverflow.com/questions/74792689}{Q74792689}: ...Every time a button is clicked, it should change its color, while the other buttons should remain the same ...So magically Copilot suggested this, and it works ...} \\
    \hline
    Web Backend & 15.90  & Discussion of server back-end and logical flow design for web application development. & \textit{\href{https://stackoverflow.com/questions/75778070}{A75778070}: ...with some guidence from ChatGPT I found this  which fixes exactly what I needed. ... chose the REST API route to point to the frontend and not the backend system ... }\\
    \hline
    Data Management & 13.78  & Discussions related to database management and data manipulation. & \textit{\href{https://stackoverflow.com/questions/76614479}{Q76614479}: I only wish to have one additional column with the information under 'dataSet''id'  ...  asked Chatgpt, it gave me this code which looks correct but just generate Traceback} \\
    \hline
    Programming script & 8.82  & Discussion of programming scripts that implement automation or other functions. & \textit{\href{https://stackoverflow.com/questions/76773098}{Q76773098}: ...I want to try out Excel Scripts. Being less familiar with JavaScript, I thought I'd try ChatGPT to do the heavy lifting...} \\
    \hline
    Algorithm & 6.54  & Discussion of algorithm design or implementation. & \textit{\href{https://stackoverflow.com/questions/76739321}{Q76739321}: I am running a simulation based on moving mass involving rayleigh beam system ... The following boundaries to be used areI have tried running this in jupyter notebook with the help of chat gpt ...} \\
    \hline
    Game  & 2.30  & Discussion of game development. & \textit{ \href{https://stackoverflow.com/questions/76701048}{Q76701048}: I'm new to coding and the game developing scene, so for my first project, I'm trying to make a Mario esc platformer in Godot. ... This occurred after asking GPT to help me add friction. This is what the code looks like now...} \\
    \hline
    Hardware & 1.59  & Discussion of hardware or embedded programming. & \textit{\href{https://stackoverflow.com/questions/76205204}{Q76205204}: I'm trying to send AT commands to HC05 for get other device names and rssi values.I'm using MSP-EXP430G2ET and HC05 module ... I asked Chat GPT to write this ...} \\
    \hline
    Others & 1.41  & Discussions that are too specific or niche to categorize. & \textit{\href{https://stackoverflow.com/questions/73684154}{Q73684154}: ...I used GitHub Copilot to help me write this next part, as I know I need to use some math to get a frequency out of this...} \\
    \hline
    \end{tabular}%
    }
  \label{tab:RQ1-LLM}%
\end{table*}%

\subsection{Answer to RQ1: What application domains do discussions on traditional LCP and LLM-based LCP focus on?}
\label{sec:RQ1}

We applied the constant comparison method and we \added{obtained} 470 instances related to the traditional LCP, distilled into four application scenario categories. We also obtained 566 instances related to the LLM-based LCP, categorized into 9 distinct application scenario categories. 
The distribution of the discussions on traditional LCP  and LLM-based LCP in various application \added{scenarios was}  presented in Tables \ref{tab:RQ1-Traditional}  and \ref{tab:RQ1-LLM}, respectively.

\noindent{\bf Application Domains.}
Table \ref{tab:RQ1-Traditional} shows that the discussions of the \added{traditional} LCP \added{have} only four categories. They focus on API integration, data management, web backend and web customization. In contrast, Table~\ref{tab:RQ1-LLM} shows that the distribution of discussions regarding LLM-based LCP \added{are spread across} 9 distinct categories. Notably, the predominant areas of focus include general issues, web frontend, web backend, and data management, which collectively account for 79.32\% of the discussions.
Additionally, this table also highlights the varied nature of LLM-based LCP applications in other domains, such as algorithms, gaming, and programming scripts.
This analysis shows that traditional LCP is more focused on application domains related to web development, whereas LLM-based LCP exhibits a broader range of application domains.

\begin{insight}
\textit{\textbf{Finding 1:} LLM-based LCP encompasses a wider array of application domains compared to traditional LCP, which are more narrowly focused on web development.}
\end{insight}

\noindent{\bf General Issues and API Integration.}
The top-ranked categories in Table \ref{tab:RQ1-Traditional} and Table \ref{tab:RQ1-LLM} highlight the prevalent topics in traditional  \added{LCP} and LLM-based \added{LCP}.
General issues are a significant focus in LLM-based LCP discussions, accounting for 32.86\%, whereas API integration takes the lead in traditional LCP discussions, reaching a substantial 36.60\%. This \added{sheds} light on the distinct challenges within these two LCPs.

In the category of general issues, users discussed issues they may face in various application areas.
For example, \added{they addressed} issues such as regular expressions, character format conversion functions, and misspelled variable names. These issues typically involve basic programming skills and syntax usage.
This means that in LLM-based LCP, users often use LLMs to generate code to solve these basic general problems, which may be difficult for novice developers to solve in manual programming.
In the category of API integration, users discussed various API integration issues they encountered when using traditional LCP tools.
For example, in \href{https://stackoverflow.com/questions/63364119}{Q63364119}, the user encountered an error when trying to add a custom button for calling an external API.
Since the underlying code in traditional LCP tools is all encapsulated into APIs, the most common problems users face often shift from generic programming problems to API integration problems.

In summary, the main discussions on traditional LCP and LLM-based LCP are related to general usage issues. The issues faced by LLM-based LCP manifest as general programming issues, while the issues faced by traditional LCP manifest as API integration issues. Both LCPs can prevent users from getting bogged down in the details of low-level code, with LLM generates professional code and traditional LCP encapsulates expert code within API.

\begin{insight}
\textit{\textbf{Finding 2:} The main discussions about traditional LCP  and LLM-based LCP revolve around general usage issues. The former predominantly involves API integration issues, whereas the latter centers on  general programming challenges.}
\end{insight}

\noindent{\bf Main Application Domains.}
In the analysis of traditional  and LLM-based LCP LCP discussions, a notable observation is the top four categories remained consistent across both contexts. These encompassed web frontend, web backend, and data management, signifying their universal importance in the LCP application domain. 

Many discussions related to traditional LCP usage involve developers seeking help on SO in areas where they encounter problems while using LCP tools. For example, \added{developers sought assistance} due to a lack of response when clicking buttons (\href{https://stackoverflow.com/questions/62646468}{Q62646468}) or inquiring about how to filter databases in the LCP tool (\href{https://stackoverflow.com/questions/63319755}{Q63319755}).
Most discussions related to LLM-based LCP are about how to use \added{LLMs} to assist in the development of these areas. For example, using Copilot to implement the functionality of changing string colors with a button click (\href{https://stackoverflow.com/questions/74792689}{Q74792689}), or using ChatGPT to generate code for manipulating JSON files (\href{https://stackoverflow.com/questions/76614479}{Q76614479}). Typically, beginners seek the help of \added{LLMs} to generate solutions when they want to implement a feature but \added{did not} know where to start.

When comparing the total proportion of these categories in Table \ref{tab:RQ1-Traditional} versus Table \ref{tab:RQ1-LLM}, we found that they accounted for 63.40\% in the discussions about traditional LCP and  46.46\% in the discussions about LLM-based LCP. This revealed that both LLM-based LCP and traditional LCP users often leverage them to accomplish tasks related to web development and data management.
This proportion is basically in line with the proportion of developers in reality. According to JetBrains' report in 2022 \cite{jetbrainsReport}, 75\% of developers \added{were} involved in web development, while developers in other fields are relatively fewer.
The proportion of traditional LCP \added{was} higher than that of LLM-based LCP, due to the fact that traditional LCP tools are usually dedicated to quickly building web applications and do not involve other domains such as algorithms and games.

\begin{insight}
\textit{\textbf{Finding 3:} Users predominantly focus on similar application areas regardless of traditional LCP  or the LLM-based LCP, i.e., web frontend, web backend and data management. }
\end{insight}

\begin{table*}[t]
  \centering
  \caption{Distribution of the discussion on traditional LCP through the software development life cycle.}
  \resizebox{1.0\textwidth}{!}{
    \begin{tabular}{m{3cm}<{\centering}|m{2.2cm}<{\centering}|m{6cm}|m{8cm}}
    \hline
    \textbf{Life cycle} & \textbf{Proportion(\%)}  & \qquad \quad \textbf{Description}  & \qquad \qquad \qquad \qquad \qquad  \textbf{Example} \\
    \hline
    Implementation & 94.99   &  Take place actual coding and development of the software. & \textit{\href{https://stackoverflow.com/questions/67962911}{A67962911}: While the expression type for a  defaults to int32, not String, it can be easily changed by opening the Properties tab in UiPath Studio and clicking the drop-down box associated with the TypeArgument attribute.} \\
    \hline
    Testing & 2.11   &  Verify and validate the software to ensure it meets the requirements and is free of bugs. & \textit{\href{https://stackoverflow.com/questions/67406513}{Q67406513}: I'm currently using cypress to test salesforce, and I'm running into a certain circumstance where I don't know the Party record ID that will create it within the opportunity...} \\
    \hline
    Deployment & 1.06   &  Release the software and made available for use, often involving installation and configuration processes. & \textit{\href{https://stackoverflow.com/questions/73917687}{Q73917687}: This code collects data from Airtable, and creates pdf file to download.It works perfectly on my localhost.But when I deploy it to Heroku server, it throws an error...} \\
    \hline
    Design & 0.79 &  Outline the software's architecture and design, and detail how it will meet the requirements. & \textit{\href{https://stackoverflow.com/questions/75826438}{Q75826438}: the webserver contacts a platform, which is used for hosting (low-code platforms such as OutSystems or CAMUNDA are closest to what I mean here), in order to run a particular Service A. This communication is managed by Event Management so that the Webserver never gets directly in contact with the service.Now, I want to represent my application as a single white-boxed component and create a context view...} \\
    \hline
    Maintain & 0.79   &  Update the software, fix any issues that arise, and potentially add new features or functionality. & \textit{\href{https://stackoverflow.com/questions/68092704}{Q68092704}: ...Google has changed something in recent weeks relating to 'caching of data in Google Sheets'. I have tried reaching out to Google Workplace support to raise this issue but they don't seem to care, and they don't offer any support for GAS... }\\
    \hline
    Project Planning \& Requirements Engineering & 0.26   &  Understand and document the client's needs and planning the overall strategy for the software development process. & \textit{\href{https://stackoverflow.com/questions/76176735}{Q76176735}: I want a password protected repository that compiles as a website for documentation, orientation, and teaching for a team of student interns... Does anyone have a suggestion about an open source solution to this?...} \\
    \hline
    \end{tabular}%
    }
  \label{tab:RQ3-traditional}%
\end{table*}%

\begin{table*}[t]
  \centering
  \caption{Distribution of the discussion on LLM-based LCP through the software development life cycle.}
  \resizebox{1.0\textwidth}{!}{
    \begin{tabular}{m{3cm}<{\centering}|m{2.2cm}<{\centering}|m{6cm}|m{8cm}}
    \hline
    \textbf{Life cycle} & \textbf{Proportion(\%)}  & \qquad \quad \textbf{Description}  & \qquad \qquad \qquad \qquad \qquad \textbf{Example} \\
    \hline
    Implementation & 87.66  &  Take place actual coding and development of the software. & \textit{\href{https://stackoverflow.com/questions/76132989}{A76132989}: Nevermind, ChatGPT came to the rescue.  The solution turned out to be using a combination of a  inside a  (this allows flutter to optimize the individual slivers) and the  package for adding the headers (using , the aformentioned  doesn't support slivers as far as I'm aware).}   \\
    \hline
    Deployment & 7.87  &  Release the software and made available for use, often involving installation and configuration processes. & \textit{\href{https://stackoverflow.com/questions/75503643}{Q75503643}: I am upgrading my app from Rails 6.1 to Rails 7.   While testing it on development env.   , it works fine.   But I encouter an issue while uploading it on heroku.   Heroku raise this error while precompiling assets :I've been trying many solutions from stackoverflow and ChatGPT...} \\
    \hline
    Design & 1.79  &  Outline the software's architecture and design, and detail how it will meet the requirements. & \textit{\href{https://stackoverflow.com/questions/76154923}{Q76154923}: I need to model a computation task and some sub-tasks depend on it... ChatGPT suggests that I could define this kind of structure as fixed point, so that I can make use of cata to fold it...} \\
    \hline
    Testing & 1.43   &  Verify and validate the software to ensure it meets the requirements and is free of bugs. & \textit{\href{https://stackoverflow.com/questions/75776530}{Q75776530}: I'm trying to write tests for an Arduino program, using VSCode with Platformio...So ChatGPT gave me a hint of using a mock digital writebut then told me to  which would work fine in Python, but not in C++. }\\
    \hline
    Maintain & 0.89  &  Update the software, fix any issues that arise, and potentially add new features or functionality. & \textit{\href{https://stackoverflow.com/questions/74959869}{Q74959869}: If I merge custom to company\_b (and it just so happens that I don't get a merge conflict) I lose the changes made in company\_b.  How can I deal with this?  According to ChatGPT I could dobut , who knows. I tried just in case and the changes I made in company\_b, specifically deleting a part of the code, are missing and the code I deleted is back.} \\
    \hline
    Requirement Analysis \& Planning & 0.36  &  Understand and document the client's needs and planning the overall strategy for the software development process. & \textit{\href{https://stackoverflow.com/questions/74962245}{Q74962245}:  As far as I know, If I want to create a new database, normally fistly I have to create and switch on a server ... I'm using ChatGPT to help me realized about any variables I don't know (Although it sometimes gives wrong information, so I try to contrast it)...} \\
    \hline
    \end{tabular}%
    }
  \label{tab:RQ3-LLM}%
\end{table*}%

\vspace{-3 mm}
\subsection{Answer to RQ2: To which software development tasks do traditional LCP and LLM-based LCP contribute?}

We identified and analyzed a total of 379 instances related to traditional LCP and a total of 559 instances associated with LLM-based LCP.
We categorized these instances \added{into} the corresponding stages of the software development life cycle (requirement analysis \& planning, design, implementation, testing, deployment, and maintenance), resulting in Table \ref{tab:RQ3-LLM} and Table \ref{tab:RQ3-traditional}. Discussions related to these two LCPs spanned the entire software development life cycle tasks.

\noindent{\bf Implementation Tasks.}
By observing Tables \ref{tab:RQ3-LLM} and \ref{tab:RQ3-traditional}, it can be seen that the implementation phase dominated in both types of discussions. Traditional LCP accounted for 94.99\%, while LLM-based LCP accounted for 87.66\%.
This phase involved actual application development, where traditional LCP typically included tasks such as customizing the user interface, implementing business logic, and debugging the implemented functionality. For example, in \href{https://stackoverflow.com/questions/67962911}{A67962911}, users discussed how to change the expression type from \textit{int32} to \textit{String} during the visual development process on the UiPath Studio platform, which is part of implementing business logic.
Tasks completed during this phase with LLM-based LCP were more extensive. In addition to the mentioned tasks, they also encompassed algorithm debugging, script writing, hardware development, and more. For instance, in \href{https://stackoverflow.com/questions/76040047}{A76040047}, the user utilized GPT4 for debugging and identifying errors occurring in encoding RSA keys.

The concentration of discussions on implementation tasks in Tables \ref{tab:RQ3-LLM} and \ref{tab:RQ3-traditional} is likely due to Stack Overflow's nature as a technical Q\&A forum that actively fostered conversations around programming techniques. Developers frequented the platform to delve into issues related to the practical aspects of coding, making the implementation phase the focal point of discussions.

\begin{insight}
\textit{\textbf{Finding 4:} Discussion posts on usage of both traditional LCP and LLM-based LCP  are prevalent across the software development life cycle, with a particular emphasis on the implementation-related tasks.}
\end{insight}

\noindent{\bf Deployment Tasks.}
A distinction \added{was observed} in deployment phase discussions between traditional LCP (1.06\%)  and LLM-based LCP (7.87\%). 
This indicates that compared with traditional LCP users, LLM-based LCP users were more concerned about deployment related topics in the software development life cycle.
This disparity aligns with the intrinsic differences between the two types of LCP.  Traditional LCP typically integrated seamlessly with cloud platforms, streamlining deployment operations through automation. This automation spared users the need to delve into the intricate details of the deployment process, creating a more user-friendly experience.
On the contrary, a feature of LLM-based LCP is in its ability to generate code that still requires a manual configuration process \added{during the deployment phase}.

\begin{insight}
\textit{\textbf{Finding 5:} LLM-based LCP discussions show a heightened focus on deployment-related tasks  compared to traditional LCP, reflecting the different needs and concerns of users in this phase of the software development life cycle.}
\end{insight}

\vspace{-3 mm}
\subsection{Answer to RQ3: What are the limitations of traditional LCP and LLM-based LCP?}

Our analysis \added{identified} a total of 95 instances related to traditional LCP and organized them into 4 categories (Table \ref{tab:RQ2-traditional}). \added{We also identified } a total of 65 instances pertaining to LLM-based LCP, which were subsequently organized into 7 distinct categories (Table \ref{tab:RQ2-LLM}).

\noindent{\bf Reliability Concerns.}
In Table \ref{tab:RQ2-LLM}, the predominant limitation observed was the reliability concern, constituting 33.33\% of the instances. 
Users expressed concerns about the reliability of generated code, as illustrated in the case of \href{https://stackoverflow.com/questions/69918631}{Q69918631}. 
Despite \added{GitHub} Copilot generating code that adhered to user instructions, concerns about the code's reliability and potential errors persisted. 
Due to users lacking sufficient professional knowledge and \added{LLMs} lacking adequate reliability, users \added{were} unable to assess the correctness and quality of the code and \added{were} concerned about the possibility of causing significant losses.

Similarly, Table \ref{tab:RQ2-traditional} shows that users of traditional LCP also exhibited concerns about reliability, at a slightly lower rate of 15.78\%.
Users' doubts about the reliability of traditional LCP \added{were} usually related to the platform's APIs or components. If these components \added{had} quality or security issues, the stability and reliability of the developed applications \added{would} be affected.
For example, in \href{https://stackoverflow.com/questions/63836762}{A63836762}, the user mentioned a kind of unreliability: since Salesforce's CometD protocol implementation \added{did} not support acknowledgment (ack), there\added{ was} a risk that subscribers will not receive messages. Users needed to design their solutions to locate and replay events that \added{had} not been committed to the target database.

The data related to reliability concerns in Tables \ref{tab:RQ2-LLM} and \ref{tab:RQ2-traditional} indicated that both LLM-based LCP and traditional LCP, especially the former, frequently \added{encountered} reliability issues. The concerns for LLM-based LCP \added{stemmed} from the inherent uncertainties and limitations of LLMs, while those for traditional LCP \added{were related} to the quality of components and potential issues with external APIs.

\begin{table*}[htbp]
  \centering
  \caption{Limitations of traditional LCP Discussed on SO}
  \resizebox{1.0\textwidth}{!}{
    \begin{tabular}{m{3cm}<{\centering}|m{2.2cm}<{\centering}|m{6cm}|m{8cm}}
    \hline
    \textbf{Limitation} & \textbf{Proportion(\%)} & \qquad \quad \textbf{Description}  & \qquad \qquad \qquad \qquad \qquad  \textbf{Example} \\
    \hline
    Need for professional knowledge & 34.74  &  Users need professional programming knowledge to use LCP tools well. & \textit{\href{https://stackoverflow.com/questions/63730131}{A63730131}: if you go further and having a complex issue that can only be solved with invoking code or creating custom activities, you really need to code.} \\
    \hline
    API function & 18.95  &  Users may meet error when using API function. & \textit{\href{https://stackoverflow.com/questions/70005251}{A70005251}: Unfortunately, airtable has a fixed 5 requests per second limit for all pricing levels. There's no way to increase this limit.} \\
    \hline
    Data migration & 16.84  &  Platform closure and domain-specific languages make data migration difficult for users. & \textit{\href{https://stackoverflow.com/questions/74056413}{Q74056413}: I am relocating our Salesforce reports from Salesforce to another tool. Salesforce has SOQL and the new system has SQL so I think the simplest way to migrate is to modify the SOQL statements to SQL, rather than recreate each report using the new tool's UI.Can I do this?} \\
    \hline
    Reliability concern & 15.79  &   The components or APIs of the LCP tool is unreliable. & \textit{\href{https://stackoverflow.com/questions/63836762}{A63836762}: Salesforce's implementation of CometD doesn't support ACKs. Even if it did, you'd still have ...but the frequency/loss of risk might be lower.In your case you have to engineer a solution that amounts to finding and replaying events that were not committed to your target database.} \\
    \hline
    Version problem & 13.68  &  The LCP tool may cause errors due to library or API version issues.  & \textit{\href{https://stackoverflow.com/questions/67410760}{A67410760}: many orgs continue using obsolete pages and components. However, public Apex classes and public Visualforce components are deleted as part of the upgrade process. If you delete pages and components without performing this two-stage procedure, Salesforce can't warn you when later deletions of public classes and components break your subscribers obsolete pages and components Emphasis mine.} \\
    \hline
    \end{tabular}%
    }
  \label{tab:RQ2-traditional}%
\end{table*}%

\begin{table*}[htbp]
  \centering
  \caption{Limitations of LLM-based LCP Discussed on SO}
  \resizebox{1.0\textwidth}{!}{
    \begin{tabular}{m{3cm}<{\centering}|m{2.2cm}<{\centering}|m{6cm}|m{8cm}}
    \hline
    \textbf{Limitation} & \textbf{Proportion(\%)} & \qquad \quad \textbf{Description} & \qquad \qquad \qquad \qquad \qquad  \textbf{Example} \\
    \hline
    Reliability concern & 33.33  &  The code generated by the LCP tool is unreliable. & \textit{\href{https://stackoverflow.com/questions/69918631}{Q69918631}: The GitHub Copilot has suggested I should use the following algorithm:So, I've tried with some testing dates and, the result is always correct. However, I'd like to know whether if this is a good way to make date comparisons or it may lead to a wrong response at any moment?} \\
    \hline
    Version Problem & 17.39  &   The LCP tool may cause errors due to library or API version issues. & \textit{\href{https://stackoverflow.com/questions/76189225}{Q76189225}: ...I even tried using chatGPT but the code generated is confusing and somehow outdated so its not working} \\
    \hline
    Need for professional knowledge & 15.94  &   Users need professional programming knowledge to use LCP tools well. & \textit{\href{https://stackoverflow.com/questions/76286719}{A76286719}: ...looking at your code you seem to lack a basic understanding of how you make Alloy models.} \\
    \hline
    Ability to guide LLM & 13.04  &   Users need to use a good prompt  or multiple rounds of dialogue for the LLM to generate properly usable code. & \textit{\href{https://stackoverflow.com/questions/76009417}{A76009417}: I was able to fix it with some ChatGPT assistance! I sent the error message to ChatGPT and it explained the following...I asked ChatGPT why is my nextjs code being run server-side?... so I asked ChatGPT can I make it so that an import only happens on the client side?... The first example didn't work out of the box, but with a little more questioning, I arrived at the solution.} \\
    \hline
    Hallucinate & 7.25  &   LLM generates code that references nonexistent functions or libraries. & \textit{\href{https://stackoverflow.com/questions/76714882}{Q76714882}: I asked ChatGPT about cross-platform and it suggested using NAudio.Alsa for Linux.  I think, it's hallucinating, since I don't see anything like that.} \\
    \hline
    Can't understand & 7.25  &   Although usable code is generated, the user does not understand the meaning of the code. & \textit{\href{https://stackoverflow.com/questions/75839167}{Q75839167}: ChatGPT was right. But WHY? ChatGPTs response was...I have to be honest. I did not understand what ChatGPT tried to explain me here.} \\
    \hline
    Privacy doubt & 5.80  &   Users are concerned about privacy being compromised when using LLM to generate code. & \textit{\href{https://stackoverflow.com/questions/72617988}{Q72617988}: I'm using the VS Code GitHub Copilot extension. Sometimes I edit files that contain secrets, and I don't want to accidentally send those to Microsoft/GitHub.} \\
    \hline
    \end{tabular}%
    }
  \label{tab:RQ2-LLM}%
\end{table*}%

\vspace{-1mm}
\begin{insight}
\textit{\textbf{Finding 6:} Users of both traditional and LLM-based LCP express concerns over reliability, with a higher degree of apprehension observed in LLM-based LCP due to the uncertainties associated with LLMs.}
\end{insight}
\vspace{-1mm}

\noindent{\bf Need for Professional Knowledge.}
LCPs aim to simplify programming tasks for professional developers and empower amateur developers with  limited coding experience. The intention is to facilitate the implementation of various functions without delving into the intricacies of traditional coding.
However, despite the advancements in LCP, the findings presented in Tables \ref{tab:RQ2-traditional} and \ref{tab:RQ2-LLM} shed light on the fact that users still require professional programming knowledge and skills, particularly when dealing with more complex functionalities.

Table \ref{tab:RQ2-traditional} and Table \ref{tab:RQ2-LLM} show that the percentage of instances expressing the ``Need for professional knowledge'' \added{amounted to} 15.94\% and 34.74\%, respectively. Notably, this statistic \added{represented} the highest proportion in Table \ref{tab:RQ2-traditional}, indicating a significant reliance on programming knowledge in traditional LCP.
As exemplified in \href{https://stackoverflow.com/questions/76286719}{A76286719}, the user pointed out that utilizing ChatGPT for constructing a simple Alloy model \added{was} mostly accurate.
However, the user still emphasized the necessity of having a fundamental understanding of Alloy model creation, reinforcing the idea that even with LLM-based LCP, users still \added{needed} to grasp foundational concepts.
Similarly, in \href{https://stackoverflow.com/questions/63730131}{A63730131}, the user believed that the LCP tool UiPath fundamentally \added{provided} the possibility of a no-code experience. Nevertheless, when confronted with intricate issues that demand the leveraging of code or the creation of custom activities, users still found themselves in need of a solid grasp of C\# knowledge to write the necessary code.

In summary, for users aspiring to incorporate more sophisticated features, the understanding of professional development skills remains crucial. LLM-based LCP users needed to understand basic programming concepts and knowledge to guide LLM in generating practical code and \added{to} validate the reliability of the generated code. Meanwhile, traditional LCP users had to possess programming skills to break free from the limitations of basic components and APIs, enabling them to implement more advanced and complex functionalities.

\vspace{-1mm}
\begin{insight}
\textit{\textbf{Finding 7:} In order to implement advanced functions, traditional and LLM-based LCP users need professional programming knowledge, with traditional LCP users needing more.}
\end{insight}
\vspace{-1mm}

\noindent{\bf Version Problem.}
Another significant issue identified in Table \ref{tab:RQ2-traditional} and Table \ref{tab:RQ2-LLM} \added{was} the ``version problem'', accounting for 17.39\% and 13.68\%, respectively.

Traditional LCP encountered many version-related problems, albeit for different reasons. In traditional LCP tools, users typically relied on the APIs or components provided by the platform and its suppliers to implement custom functionalities. However, when a component underwent an upgrade and conflicts with some common components, users might find that their custom functions no longer operate as intended.
An insightful instance \added{was} highlighted in \href{https://stackoverflow.com/questions/76189225}{A76189225}, users in Salesforce mentioned that \added{conflicted} between ``many organizations are still using outdated pages and components'' and ``common component upgrades'' \added{resulted} in errors without warnings.

In LLM-based LCP, version issues \added{arose} due to the static nature of LLM training data, which may not align with current library updates and programming environments. As a result, the generated code \added{could} be inconsistent with the libraries and \added{environments} used by the user. Prompting issues might produce less than accurate LCP code. Additionally, the hallucination of LLM might lead to incorrect versions of functions. These issues resulted in version-related limitations for users of LLM-based LCP.
An illustrative example of this limitation could be found in \href{https://stackoverflow.com/questions/76189225}{Q76189225}, where the user attempted to implement certain functions with the assistance of code generated by ChatGPT. However, the generated code turned out to be outdated.

Overall, version issues faced by LLM-based LCP \added{tended} to be more critical, often leading to project compilation failures. In contrast, traditional LCP's version issues primarily \added{arose} during the maintenance phase, requiring providers to promptly maintain components and resolve version conflicts.

\begin{insight}
\textit{\textbf{Finding 8:} 
Both traditional LCP and LLM-based LCP users face version-related issues, with the former due to conflicts during component upgrades, and the latter stemming from the outdated data and hallucinate of LLM.
} 
\end{insight}

\noindent{\bf API Function.}
In Table \ref{tab:RQ2-traditional}, it was noteworthy that API function related discussions constituted a substantial portion, ranking the second position at 18.95\%.
This underscored a common challenge faced by traditional LCP that heavily \added{relied} on pre-defined APIs, where users encountered functional constraints due to the inherent limitations of these interfaces.
For example, in \href{https://stackoverflow.com/questions/70005251}{A70005251}, the user \added{encountered} API performance limitations, which might \added{posed} a bottleneck for developers seeking to implement high-level customization of functionality within their applications.

Compared to traditional LCP, LLM-based LCP significantly deviated from these limitations. 
LLM-based LCP \added{had} strong learning capabilities and the ability to autonomously generate code. Its powerful learning ability \added{enabled} it to adapt to constantly changing user requirements, no longer constrained by fixed API interfaces, thus handling more complex scenarios. In contrast, traditional LCP \added{relied} on static APIs and \added{could not} proactively adapt to changing user needs, resulting in a noticeable disadvantage in flexibility.
Furthermore, when implementing a customized functionality, LLM-based LCP \added{could} provide solutions and automatically \added{generated} code. In contrast, traditional LCP often \added{required} developers to manually write a large amount of code when dealing with functional customization, increasing the development workload.

\vspace{-1mm}
\begin{insight}
\textit{\textbf{Finding 9: }LLM-based LCP appears to surpass traditional LCP's API limitations, enhancing flexibility and adaptability.}
\end{insight}
\vspace{-1mm}

\noindent{\bf Data Migration.}
Data migration also played a high role in the discussion of traditional LCP, reaching 16.84\%.
Data migration referred to the fact that users might want to migrate data to another LCP platform because of API limitations, functional limitations, etc. However, due to domain-specific languages (such as the specific query languages for each platform), migrating data or functionality from one platform to another became quite complex.

For example, in \href{https://stackoverflow.com/questions/74056413}{Q74056413}, the user attempted to migrate data and reports from the sales platform to another platform. But Salesforce had SOQL (salesforce object query language), and the new platform also had specific SQL. The user encountered problems during the data migration process and did not know how to convert a large number of SOQL statements.

For traditional LCP, users \added{needed} to carefully choose a platform for their applications because each platform \added{had} its own domain-specific language and unique APIs. Once data and functionality \added{were} tightly integrated on a particular platform, migrating to another platform \added{could} require a complete application refactor, resulting in significant costs. This \added{meant} that users when selecting an LCP \added{tool} \added{needed} to balance not only current requirements but also consider potential changes and expansions in the future.
In contrast, LLM-based LCP typically \added{provided} developers with more flexibility and control. Developers \added{could} manage data more freely, \added{were} no longer constrained by a specific platform; and thus, they \added{did} not need to worry about data migration issues between LCP platforms.

\begin{insight}
\textit{\textbf{Finding 10:} Traditional LCP poses data migration challenges due to its closed nature, while LLM-based LCP offers greater flexibility in this regard.}
\end{insight}

\noindent{\bf Uncertainty of LLM-based LCP.}
Users of LLM-based LCP faced limitations on ``version problem'', ``ability to guide LLM'' and ``hallucinations'', which reached 17.39\%, 13.04\% and 7.25\%, respectively. 
These issues underscore the inherent uncertainty in using LLM-based LCP for programming and \added{that} software development \added{did} not guarantee satisfactory results. There was a risk of generating erroneous code or unsatisfactory solutions due to the hallucination problems of LLM or the developers' lack of ability to guide LLM.


To understand user satisfaction with using LLM-based LCP, we conducted a detailed analysis of the data related to LLM-based LCP in~Section \ref{sec:RQ1}. In Section~\ref{sec:RQ1}, we had already extracted 566 instances related to LLM-based LCP and discussed their distribution in application domains. 
We reanalyzed the aforementioned 566 instances to determine whether users provided positive feedback. For example, in \textit{\href{https://stackoverflow.com/questions/75778070}{A75778070}}, the user mentioned, \textit{"...asked ChatGPT, it gave me this code which looks correct but just generates a Traceback"}, which \added{was} considered positive feedback. The judgment was made by the first author and verified by the second author.

The data presented in Table~\ref{tab:LLM_success_percentage} illustrates the proportion of LLM-based LCP users who provided positive feedback across various application fields. The overall positive feedback rate \added{stood} at 39.22\%. These findings suggest that users may not be highly satisfied with the solution generation using LLM-based LCP.

Overall, while LLM-based LCP offers innovative prospects, its \added{positive feedback rate in practical applications} \added{remained} modest. To enhance the stability of LLM-based LCP, developers and researchers \added{needed} to focus on and continuously improve above aspects to ensure that LLM-based LCP \added{could} better meet the needs of users.

\vspace{-1mm}
\begin{insight}
\textit{\textbf{Finding 11:} Compared to traditional LCP, LLM-based LCP exhibits greater uncertainty, necessitating improvements to meet user expectations more reliably.}
\end{insight}
\vspace{-1mm}


\begin{table}[t]
  \centering
  \caption{Positive feedback rate from LLM-based LCP users. PFR, PFI, and Ins. respectively represent positive feedback rate, positive feedback instances, and total instances.}
  \resizebox{0.4\textwidth}{!}{
    \begin{tabular}{cccc}
    \toprule
    \textbf{Application Areas} & \textbf{PFR(\%)} & \textbf{PFI} & \textbf{Ins.} \\
    \midrule
    General & 43.01 & 78 & 186  \\
    Web Frontend & 33.68 & 32 & 95  \\
    Web Backend & 23.33 & 21  & 90 \\
    Database Management & 48.72 & 38  & 78  \\
    Simple script & 36.00 & 18  & 50  \\
    Algorithm & 62.16 & 23  & 37  \\
    Game  & 23.08 & 3  & 13  \\
    Hardware & 44.44 & 4  & 9  \\
    Others & 62.50 & 5  & 8  \\
    \midrule
    Total & 39.22 & 222  & 566  \\
    \bottomrule
    \end{tabular}%
    }
  \label{tab:LLM_success_percentage}%
\end{table}%

\section{DISCUSSION}
\label{Discussions}
Although traditional LCP and LLM-based LCP \added{exhibited} different characteristics (RQ1) and limitations (RQ3), there \added{was} a noticeable trend toward convergence in their key technologies. 
We first \added{discussed} how LLM/AI technology \added{had recently been incorporated} into traditional LCP tools in Section \ref{sec:LLM_drive_LCP}. Following that, in Section \ref{sec:VPLs_drive_LLM}, we \added{explored} how visualization technology \added{was} used by recent LLM-based LCP tools. 
Section \ref{sec:Agent} \added{was} dedicated to examining the implications of `LLM Agents' for LCP. Lastly, in Section \ref{sec:threats}, we \added{discussed} the potential threats to the validity of this study.

\subsection{LLM/AI Technology Added to Traditional LCP Tools  }
\label{sec:LLM_drive_LCP}

Our analysis of website descriptions and documentation of traditional LCP tools 
presented in Section \ref{Methodology} \added{revealed} that 13 out of these 24 tools \added{had} incorporated AI technology since 2023.
These functionalities \added{were} categorized into LLM-driven features and AI-driven features. LLM-driven features \added{referred} to the use of LLM in LCP tools to generate applications, perform data analysis, or generate workflows and code. On the other hand, AI-driven features \added{referred} to features driven by non-LLM AI, such as automatically recommending components that could be applied.
By carefully reading their official documentation and experiencing the AI features of these LCP tools, we summarized several features as shown in Table \ref{tab:LLM_in_traditional_LCP}.


\begin{table*}[t]
  \centering
  \caption{Recent AI-based features added to traditional LCP tools}
  \resizebox{0.95\textwidth}{!}{
    \begin{tabular}{c|c|c|c|c}
    \hline
    \multirow{2}[4]{*}{\textbf{LCP Tools}} & \multicolumn{3}{c|}{\textbf{LLM-Driven}} & \textbf{AI-Driven} \\
\cline{2-5}          & \textbf{Generate Apps} & \textbf{Data Analysis} & \textbf{Generate Workflow/Code } & \textbf{Recommend Field/Component} \\
    \hline
    Quickbase \cite{quickbase} & \checkmark      &       &       & \checkmark  \\
    \hline
    OutSystems \cite{outsystems} & \checkmark      &   \checkmark    &   \checkmark    & \checkmark  \\
    \hline
    Power Apps  \cite{powerapps}& \checkmark      &       & \checkmark      & \checkmark  \\
    \hline
    Appypie \cite{appypie} & \checkmark      &       &       &  \\
    \hline
    Appsheet \cite{appsheet} & \checkmark      &       &       &  \\
    \hline
    Softr \cite{softr}  & \checkmark      &       &       &  \\
    \hline
    Appian  \cite{appian}&       & \checkmark      & \checkmark      &  \\
    \hline
    Servicenow \cite{servicenow} &       & \checkmark     & \checkmark     &  \\
    \hline
    Airtable  \cite{airtable}&       & \checkmark      & \checkmark      & \checkmark  \\
    \hline
    Mendix \cite{mendix}&       &       &       & \checkmark  \\
    \hline
    Uipath \cite{uipath} &       &    \checkmark   &   \checkmark    &  \\
    \hline
    Pega \cite{pega} &       &       &       & \checkmark  \\
    \hline
    Nintex \cite{nintex}&       &       &       & \checkmark  \\
    \hline
    \end{tabular}%
    }
  \label{tab:LLM_in_traditional_LCP}%
\end{table*}%

\subsubsection{App Generation Driven by LLM}
Combined with LLM technology, users can generate applications using natural language in LCP tools without any manual or even visual programming.
For example, in \textsc{Quickbase} \cite{quickbase}, users can use natural language to describe things like, ``\textit{What do users want the app to do? Or what problem are users trying to solve?} " Then, Quickbase can produce a fully functional application that meets \added{their} needs. After generating the application, users can make further detailed adjustments to it based on their requirements within the visual development interface.
In \textsc{Softr} \cite{softr}, users simply only need to select the application type and give prompt to describe the goals and functions of the application, then \textsc{Softr} can generate the application directly.
This feature significantly saves time for programming development and can save time and learning costs for beginners. However, the generated applications by this technology tend to be simple and monotonous, making them better suited for straightforward functionalities, inspiration, and quick validation of ideas.

\subsubsection{Workflow/Code Generation Driven by LLM}
Users can generate workflows or code snippets using natural language directly in the LCP tool.
Unlike generating applications using natural language mentioned earlier, this feature emphasizes more on the possibility of controllably generating or modifying parts of the application using natural language. 
For example, in \textsc{Appian} \cite{appian}, users can generate a workflow of the complete process by talking to the built-in copilot and presenting it as a visualization module. This can then be reviewed and modified in a low-code visualization model in \textsc{Appian}.
Similarly, in \textsc{Appsheet} \cite{appsheet}, developers talk to the built-in Duet AI to generate automated actions on the data. For example, ``\textit{Check if everything is working properly every week and submit a report.}'' Then, Appsheet will automatically create forms to track the data.
This feature enables a deep integration between traditional LCP and LLM-based LCP tools. Developers can not only generate applications using natural language but also exercise finer control over the generated applications on a visual canvas for meeting the specific requirements of their applications. 

\subsubsection{Data Analysis Driven by LLM}
This feature refers to the ability for developers to utilize natural language to explore, summarize, analyze data, or automatically generate reports. For example, in \textsc{Airtable} \cite{airtable}, customer service professionals can make natural language requests to AI to categorize their feedback data based on emotions such as positive, neutral, and negative. Subsequently, AI will autonomously perform this task and generate the corresponding table.
In \textsc{Appian}, users can ask questions to the built-in LLM and seek specific insights into in-app data by using natural language.
Integrating LLM into traditional LCP tools \added{empowers} them with data analysis capabilities. This feature not only allows developers to interact with data in a natural and intuitive way but also reduces the learning curve for developers who may not have a background in data science or complex query languages.

\subsubsection{Field/Component Recommendation  Driven by AI}
Through learning user needs, project characteristics, and developer preferences, LCP tools can recommend the most suitable fields or components for the current context during the development process, thereby accelerating the development process.
For example, when developing software in \textsc{OutSystems} \cite{outsystems} platform, in the visual programming interface, logic flows are constructed by dragging and dropping components. \textsc{OutSystems} will recommend to the developer the next possible components and the parameters of these components based on the current logic flow. This knowledge is derived from learning about 250,000 anonymous patterns in \textsc{OutSystems}.
This feature further speeds up the traditional LCP development workflow and reduces the cognitive load on the LCP tool for developers.

When investigating the AI-driven features added to traditional LCP tools, we observe an apparent trend that has emerged. The progressive integration of AI technologies within these platforms has served as a catalyst for heightened development efficiency and a reduction in the learning curve.
Moreover, mainstream traditional LCP tools are actively exploring how to integrate LLM within the tool to maximize the advantages of LLM-based LCP and combine them with the strengths of traditional LCP.

\subsection{Visualization drives LLM-based LCP}
\label{sec:VPLs_drive_LLM}

In the combination of LLM and visualization technology, in addition to the introduction of LLM in traditional LCP tools, there are some studies that use visualization technology to drive \added{LLM-based} LCP.
Since LLMs at this stage are still not capable of aligning the generated code with the user's prompts, some researchers are attempting to bridge the gap by leveraging various visualization techniques.

\subsubsection{LowCoder: Integrating Visual and Natural Language Interfaces for AI Pipeline Development.}

\textsc{LowCoder}~\cite{rao2023ai} is an LCP tool for developing AI pipelines that supports both a visual programming interface and an LLM-based natural language interface.
Specifically, \textsc{LowCoder} uses visual programming as a read-write view and PBNL as a writing-only view, letting developers view data in a read-only view. The tool keeps these three views in sync by representing the program in a domain-specific language (DSL). \added{Users} can drag and drop any block of operators from the palette onto the canvas to form an AI pipeline, and set hyperparameters for each operator.

However, the tool palette contains more than a hundred blocks of operators, making it very difficult for the user to find the desired operator to use. \textsc{LowCoder} therefore provides an NL interface where the user can describe the desired operation in a text box and press the ``Predict Pipeline'' button. The tool then uses a natural language code-switching model to infer the relevant operators and any applicable hyperparameters, and automatically adds the most relevant operators to the end of the pipeline.
In this tool, researchers use VPLs as the main approach to develop AI pipelines, supplemented by PBNL to view data. Additionally, employing LLM to recommend the next set of operators and parameters further accelerates development efficiency.

\subsubsection{CoLadder:  Enhanced Control Over LLM Code Generation with Visual Tools.}

\textsc{CoLadder}~\cite{yen2023coladder} uses visualization to give developers control over the LLM code generation process.
Specifically, \textsc{CoLadder} contains a tree-based prompt editor that allows developers to split tasks into manageable subtasks.
This allows for more flexible construction of hierarchical prompt structures, resulting in aligned code structures that are not only easy to generate but also straightforward to verify.

In addition, \added{the} tool supports direct manipulation of the prompt structure through various prompt block operations. Developers can seamlessly \added{add, edit, delete, or drag and drop} prompt blocks via intuitive buttons. Each block-based operation triggers updates to the corresponding code and ensures that changes are propagated consistently to the rest of the code as needed.
Finally, \textsc{CoLadder} utilizes a block-based design to modularize each prompt and snippet of code. After each module's prompt generates the code, the code can be modified flexibly in the block editor.

Overall, the design of \textsc{CoLadder} builds upon LLM-based LCP, supplemented with VPLs. This facilitates code generation and validation, providing developers with a high level of control and intuitiveness, thereby further reducing the threshold for developers to guide LLM.

\subsubsection{Low-code LLM: Transforming Visual Workflows and Natural Language Descriptions.}


Low-code LLM~\cite{cai2023low} is \added{an} LLM tool with visual interfaces for a wide variety of tasks, including programming.
The user can first enter a short task description, and LLM will complete a prompted workflow in a specific format (with step-by-step descriptions and logical jumps) based on the task description, and visualize it to the user on the canvas.
The canvas serves as an interactive space where developers have the flexibility to refine and customize the generated workflow through six distinct actions: add/remove steps, modify descriptions, add/remove a jump logic, drag and drop, extend sub-flowchart, and regenerate and confirm.
Finally, after getting confirmation, the tool translates the specific format workflow into a natural language description and generates the code for the entire workflow by the LLM.

Unlike the previous two tools, this tool places more emphasis on using a specific format to transform visual workflows and natural language descriptions to each other, and directly inputs all generated natural language descriptions into LLM to generate code

\subsection{LLM Agent for LCP}
\label{sec:Agent}

In Section \ref{sec:Results}, we found that developers primarily use LLM-based LCP during the implementation phase of development (Finding 4). Due to the \added{hallucinations} and uncertainties associated with \added{LLMs}, developers require specialized programming knowledge to design programming architectures or ensure code correctness (Finding 7). Some studies \added{attempted} to bridge this gap by employing agents based on LLM \cite{hong2023metagpt,autoGPT,qian2023communicative}. These agents \added{could} span the entire software development life cycle, reducing logical inconsistencies and errors in LLM. The following is a brief introduction to the three LLM agents to understand the impact of the agent on the LCP.

\subsubsection{AutoGPT: LLM Single-Agent for Software Development }
\textsc{AutoGPT}~\cite{autoGPT} is the most famous GPT agent project that can be used in software development, business analysis, market research, and other fields. As of the time of writing this paper, there are 155k stars on GitHub.
Developers only need to provide a prompt or a set of natural language instructions. \textsc{AutoGPT} decomposes the target into subtasks, which are then linked together and executed in order to produce larger results initially arranged by user input.
In the field of software engineering, \textsc{AutoGPT} can assign different roles to intelligent agents, thereby autonomously executing the capabilities of different stages in the software development life cycle. However, due to the fact that \textsc{AutoGPT} did not optimize collaboration between different agents, this tool still has hallucinations and is difficult to solve complex engineering problems.

\subsubsection{MetaGPT: LLM Multi-Agent for Software Development }
In MetaGPT~\cite{hong2023metagpt}, five roles are defined that span the entire software development cycle: Product Manager, Architect, Project Manager, Engineer, and QA Engineer, and they are assigned features, goals, and constraints. All agents follow the software development SOP, such as the Product Manager conducting in-depth analysis of user requirements, developing product requirements documents, and then handing them over to the Architect. In addition, Engineer and QA Engineer can communicate with each other to correct any errors that may exist in the generated code. This tool effectively decomposes complex software engineering tasks into subtasks involving collaboration between multiple agents, allowing more complex software to be developed.

\subsubsection{ChatDev: LLM Multi-Agent for Software Development Based on Chat Chain}
This tool~\cite{qian2023communicative} divides the development process into four stages: design, coding, testing, and documentation. At each stage, \textsc{ChatDev} recruits multiple agents with different roles, such as programmers, reviewers, and testers.
Through the chat chain, each stage is divided into atomic subtasks, and two adjacent roles in the chat chain participate in context aware multi round discussions to propose and validate solutions.
Unlike the MetaGPT, the proposed chat chain of this tool collaborates through natural language.
This multi-agent assisted system also spans the entire software development cycle, alleviating the hallucinations of LLM and providing inspiration for future low-code development.

In summary, the concept of an LLM Agent for LCP provides a different solution than combining visualization to solve the complexity and unreliability of LLM in the software development life cycle. This is via collaborative communication among multiple agents, providing valuable insights for future development of LCP.

\subsection{Threats to Validity}
\label{sec:threats}

\textbf{Internal Validity: } Key internal threats are that authors might be biased when performing data analysis. 
To reduce subjective biases and enhance the dependability of our findings, we adopted a dual-author approach for each analysis, guaranteeing comprehensive resolution of any disagreements that arose. \added{We measured inter-coder agreement, yielding a Cohen’s kappa of 0.89, indicating strong consistency between annotators.}  

\added{In addition, when constructing the LCP dataset based on LLMs, we adopted a two-stage process of keyword pre-screening and ChatGPT semantic filtering. Since the second stage may introduce bias, we carefully designed and standardized the prompts. This design was informed by prior work on prompt engineering to ensure a higher recall rate (86.7\%). To control potential false positives resulting from the high recall, all filtered posts underwent a second round of manual verification to ensure overall data quality. At the same time, we acknowledge the non-determinism of large language models; in future work, we plan to explore more robust approaches, such as model fine-tuning and ensemble/multi-agent frameworks, to further enhance reproducibility and stability.}

\noindent \textbf{External Validity: }
Key external threats are related to the generalizability of our research findings. Our study is based on discussion data from developers on SO. However, since LLM-based LCP is a relatively recent development and is still evolving and changing rapidly, discussions on SO may not necessarily represent the latest state of affairs. Additionally, while SO is a leading platform for developer Q\&A, it may not fully represent the diverse discussions occurring in the broader programming community. 
\added{In particular, discussions of traditional LCP on SO may overlook its adoption in enterprise or non-developer contexts, where such platforms are widely used but less often discussed in public forums.
To ensure high-quality discussions, we manually evaluated data by selecting posts based on their scores, from high to low. 
We explicitly acknowledge this single-source reliance as a threat to validity. While this study focuses on Stack Overflow data, future work will extend the analysis by triangulating these findings with supplementary sources such as technical blogs, and GitHub discussions.}

\section{Related Work}
\label{related}


\vspace{-1mm}
\subsection{Low-Code Programming}

With the growing popularity of low-code approaches, LCP has been applied in various fields such as human resources, the Internet of Things, machine learning, etc \cite{di2020democratizing,ihirwe2020low,oteyo2021building,martins2020overview}.
Zhuang et al. \cite{zhuang2022easyfl} proposed a low-code federated learning framework. By simplifying APIs and adopting a modular design, this framework enables users with varying levels of expertise to experiment with federated learning applications with minimal coding.
Chen et al. \cite{chen2022devicetalk} proposed DeviceTalk, a low-code IoT development framework that utilizes visual techniques to accelerate software development in the field of distributed intelligent system devices.

Because the concept of low-code is relatively vague, some studies have explored the concept and models of low-code. 
Di Ruscio et al. \cite{di2022low} compared low-code with model-driven approaches, identified their similarities and differences, and analyzed the strengths and weaknesses of these two methods. 
Gomes and Brito \cite{gomes2022low} conducted a descriptive study on low-code, exploring and showcasing key concepts, factors, and variables related to low-code development platforms.
Hirzel \cite{hirzel2023low}  integrated low-code literature from various research fields, explained the technical principles of low-code programming, and provided a unified perspective on it.

In addition, some studies have empirically investigated LCP from the perspective of developers. 
Rafi et al. \cite{rafi2022devops} conducted interviews with developers using low-code to understand practitioners' views on the low-code trend. This research found that the emergence of low-code can greatly facilitate the development of high-quality products with low cost and time.
Luo et al. \cite{luo2021characteristics}  used data from SO and Reddit to analyze the descriptions, advantages, limitations, and challenges of low-code development.
Al Alamin et al. \cite{al2021empirical} conducted empirical research on SO data. This research used topic modeling to analyze the distribution of discussion topics on SO, thus identifying the difficulties users face in the process of using LCP.

There are some studies that summarize the characteristics and challenges of LCP through literature reviews.
Sufi \cite{sufi2023algorithms} conducted an extensive literature review on topics such as low-code, no-code, visual programming, and model-driven programming. The review explored the advantages, limitations, features, and application areas of LCP.
Rokis and Kirikova \cite{rokis2022challenges} carried out a literature review to systematically summarize various challenges in low-code software development.

Although the above-mentioned papers have explored the concepts, characteristics, and challenges of LCP using various methods, these studies have confined the concept of LCP solely to tools related to visualization and modularity, namely the traditional LCP proposed in this paper. The characteristics, application domains, and limitations faced by LLM-based LCP are distinct from traditional LCP. Thus, the latest LLM-based LCP requires further research.

\vspace{-1mm}
\subsection{Large Language Models for Software Engineering}
LLM have garnered considerable attention in academic and industrial circles, showcasing remarkable proficiency in various tasks across multiple disciplines\cite{bubeck2023sparks}.
Notably, they have the ability to generate usable code, providing invaluable support to LCP \cite{chen2021evaluating,ross2023programmer,liu2023wants}.
In the field of software development, tools such as GPT4 and Copilot can generate code based on the user's instructions \cite{kazemitabaar2023studying,peng2023impact,nguyen2022empirical,vaithilingam2022expectation}. Users can talk to a LLM, describe their task, and let it generate code to complete the task. Alternatively, give the starting code snippet and let it continue. LLM increases the efficiency of software developers and helps amateur developers develop.

Several studies investigated the code generation ability of LLM and its applications in the field of software engineering.
Liang et al. \cite{liang2023large} surveyed the users of AI Programming Assistants to understand how developers use these tools and the usability challenges they face.

Sridhara et al. \cite{sridhara2023chatgpt} examined the utilization of ChatGPT in executing various common software engineering tasks and compared and analyzed using ground facts from human experts.
Poldrack et al. \cite{poldrack2023ai} studied the ability to complete programming tasks and the quality of generated code using GPT4.
Bubeck et al. \cite{bubeck2023sparks} demonstrated in a set of coding challenges that GPT4's coding capabilities achieved human-level performance.
Suri et al. \cite{suri2023software} delved into the capabilities and challenges of auto-GPT in software engineering practices, especially dealing with complex frameworks such as Sping Boot, Django, and Flask.

The above studies investigated the programming capabilities of LLMs and agents based on LLMs, as well as their ability to perform software engineering tasks. These studies also examined the challenges faced by developers applying LLMs. However, the literature mentioned above did not explore the characteristics and limitations of LLMs from the perspective of LCP. Our study explores the similarities and differences between LLM-based LCP and traditional LCP, discussing how LLM-based LCP and traditional LCP can be integrated.

\section{Conclusion and Future work}
\label{conclusion}
In this paper, we delve into the similarities and differences between traditional LCP and LLM-based LCP. Our empirical study involved collecting related posts on SO over the past three years to determine the similarities and differences.
Our findings highlight commonalities and disparities in the development areas, limitations, and distribution of software development cycles.
They are mainly used in the field of web development and face limitations such as ``Reliablity doubt" and ``Need for professional knowledge". Their discussions are similarly distributed throughout the software development cycle.
However, LLM-based LCP applies to a wider range of scenarios and solves more problems, facing problems specific to larger language models such as "Hallucinate". In discussions of the software development cycle, there is also more discussion of deployment issues.
To the best of our knowledge, this is the first empirical study of LLM-based LCP. This research can help developers of low-code tools to better understand and integrate the two. 

\added{In future work, we will extend our analysis beyond SO by incorporating diverse data sources such as GitHub Discussions and technical blogs, and applying triangulation to mitigate single-source bias. This will enable cross-platform comparisons in terms of question types, application domains, and tool usage, thereby enhancing the robustness of our conclusions. In addition, we plan to conduct user studies, such as surveys and interviews with practitioners of low-code tools, to gather direct feedback on their experiences with LLM-based and traditional LCPs, thereby validating and refining our findings.}


\section{Conflict of Interests}

The authors declared that they have no conflict of interest exits in the submission of this manuscript, and manuscript is approved by all authors for publication. I would like to declare on behalf of my co-authors that the work described was original research that has not been published previously and is not under consideration for publication elsewhere. All the authors listed have approved the manuscript.

We declare that we do not have any commercial or associative interest that represents a conflict of interest in connection with the work submitted.

\section{Acknowledgements}
This work is supported by the National Natural Science Foundation of China (No. 62332004), the National Natural Science Foundation of China (62276279, 62302534), Guangdong Basic and Applied Basic Research Foundation (2024B1515020032, 2025A1515011632) and the National Key Research and Development Program of China (2023YFB2703700).

\section{Data availability statements}

The dataset and results of this paper are open source and available in \UrlFont{https://zenodo.org/records/11232842}.






\begin{thebibliography}{00}





\bibitem{bock2021low}
A.~C. Bock and U.~Frank, ``Low-code platform,'' \emph{Business \& Information Systems Engineering}, vol.~63, pp. 733--740, 2021.

\bibitem{hirzel2023low}
M.~Hirzel, ``Low-code programming models,'' \emph{Communications of the ACM}, vol.~66, no.~10, pp. 76--85, 2023.

\bibitem{sahay2020supporting}
A.~Sahay, A.~Indamutsa, D.~Di~Ruscio, and A.~Pierantonio, ``Supporting the understanding and comparison of low-code development platforms,'' in \emph{2020 46th Euromicro Conference on Software Engineering and Advanced Applications (SEAA)}.\hskip 1em plus 0.5em minus 0.4em\relax IEEE, 2020, pp. 171--178.

\bibitem{bragancca2021towards}
A.~Bragan{\c{c}}a, I.~Azevedo, N.~Bettencourt, C.~Morais, D.~Teixeira, and D.~Caetano, ``Towards supporting spl engineering in low-code platforms using a dsl approach,'' in \emph{Proceedings of the 20th ACM SIGPLAN International Conference on Generative Programming: Concepts and Experiences}, 2021, pp. 16--28.

\bibitem{burnett1995visual}
M.~M. Burnett and D.~W. McIntyre, ``Visual programming,'' \emph{COmputer-Los Alamitos-}, vol.~28, pp. 14--14, 1995.

\bibitem{cypher1993watch}
A.~Cypher and D.~C. Halbert, \emph{Watch what I do: programming by demonstration}.\hskip 1em plus 0.5em minus 0.4em\relax MIT press, 1993.

\bibitem{jiang2022discovering}
E.~Jiang, E.~Toh, A.~Molina, K.~Olson, C.~Kayacik, A.~Donsbach, C.~J. Cai, and M.~Terry, ``Discovering the syntax and strategies of natural language programming with generative language models,'' in \emph{Proceedings of the 2022 CHI Conference on Human Factors in Computing Systems}, 2022, pp. 1--19.

\bibitem{powerapps}
Microsoft. (2023) Power apps. [Online]. Available: \url{https://powerapps.microsoft.com}

\bibitem{uipath}
U.~Team. (2023) Uipath. [Online]. Available: \url{https://www.uipath.com}

\bibitem{brown2020language}
T.~Brown, B.~Mann, N.~Ryder, M.~Subbiah, J.~D. Kaplan, P.~Dhariwal, A.~Neelakantan, P.~Shyam, G.~Sastry, A.~Askell \emph{et~al.}, ``Language models are few-shot learners,'' \emph{Advances in neural information processing systems}, vol.~33, pp. 1877--1901, 2020.

\bibitem{poldrack2023ai}
R.~A. Poldrack, T.~Lu, and G.~Begu{\v{s}}, ``Ai-assisted coding: Experiments with gpt-4,'' \emph{arXiv preprint arXiv:2304.13187}, 2023.

\bibitem{bubeck2023sparks}
S.~Bubeck, V.~Chandrasekaran, R.~Eldan, J.~Gehrke, E.~Horvitz, E.~Kamar, P.~Lee, Y.~T. Lee, Y.~Li, S.~Lundberg \emph{et~al.}, ``Sparks of artificial general intelligence: Early experiments with gpt-4,'' \emph{arXiv preprint arXiv:2303.12712}, 2023.

\bibitem{StackOverflow}
S.~O. Team. (2023) Stack overflow. [Online]. Available: \url{https://stackoverflow.com}

\bibitem{glaser1965constant}
B.~G. Glaser, ``The constant comparative method of qualitative analysis,'' \emph{Social problems}, vol.~12, no.~4, pp. 436--445, 1965.

\bibitem{F2014}
R.~Clay, R.~John, R, M.~Christopher, C.~Alex, and W.~Dominique, ``Vendor landscape: A fork in the road for low-code development platforms,'' \emph{Analyst Report. Forrester Research Inc}, 2014.

\bibitem{richardson2016forrester}
C.~Richardson and J.~R. Rymer, ``The forrester wave™: low-code development platforms, q2 2016,'' \emph{Forrester, Washington DC}, 2016.

\bibitem{rymer2017vendor}
J.~R. Rymer \emph{et~al.}, ``Vendor landscape: A fork in the road for low-code development platforms,'' \emph{Analyst Report. Forrester Research Inc}, 2017.

\bibitem{di2022low}
D.~Di~Ruscio, D.~Kolovos, J.~de~Lara, A.~Pierantonio, M.~Tisi, and M.~Wimmer, ``Low-code development and model-driven engineering: Two sides of the same coin?'' \emph{Software and Systems Modeling}, vol.~21, no.~2, pp. 437--446, 2022.

\bibitem{van2018robotic}
W.~M. Van~der Aalst, M.~Bichler, and A.~Heinzl, ``Robotic process automation,'' pp. 269--272, 2018.

\bibitem{outsystems}
O.~Team. (2023) Outsystems. [Online]. Available: \url{https://www.outsystems.com/}

\bibitem{copilot}
Github. (2023) Github copilot. [Online]. Available: \url{https://github.com/features/copilot}

\bibitem{chatgpt}
OpenAI. (2023) Chatgpt. [Online]. Available: \url{https://openai.com/chatgpt}

\bibitem{vaswani2017attention}
A.~Vaswani, N.~Shazeer, N.~Parmar, J.~Uszkoreit, L.~Jones, A.~N. Gomez, {\L}.~Kaiser, and I.~Polosukhin, ``Attention is all you need,'' \emph{Advances in neural information processing systems}, vol.~30, 2017.

\bibitem{hou2023large}
X.~Hou, Y.~Zhao, Y.~Liu, Z.~Yang, K.~Wang, L.~Li, X.~Luo, D.~Lo, J.~Grundy, and H.~Wang, ``Large language models for software engineering: A systematic literature review,'' \emph{arXiv preprint arXiv:2308.10620}, 2023.

\bibitem{ma2023scope}
W.~Ma, S.~Liu, W.~Wang, Q.~Hu, Y.~Liu, C.~Zhang, L.~Nie, and Y.~Liu, ``The scope of chatgpt in software engineering: A thorough investigation,'' \emph{arXiv preprint arXiv:2305.12138}, 2023.

\bibitem{qian2023communicative}
C.~Qian, X.~Cong, W.~Liu, C.~Yang, W.~Chen, Y.~Su, Y.~Dang, J.~Li, J.~Xu, D.~Li, Z.~Liu, and M.~Sun, ``Communicative agents for software development,'' 2023.

\bibitem{hong2023metagpt}
S.~Hong, M.~Zhuge, J.~Chen, X.~Zheng, Y.~Cheng, C.~Zhang, J.~Wang, Z.~Wang, S.~K.~S. Yau, Z.~Lin, L.~Zhou, C.~Ran, L.~Xiao, C.~Wu, and J.~Schmidhuber, ``Metagpt: Meta programming for a multi-agent collaborative framework,'' 2023.

\bibitem{zamfirescu2023johnny}
J.~Zamfirescu-Pereira, R.~Y. Wong, B.~Hartmann, and Q.~Yang, ``Why johnny can’t prompt: how non-ai experts try (and fail) to design llm prompts,'' in \emph{Proceedings of the 2023 CHI Conference on Human Factors in Computing Systems}, 2023, pp. 1--21.

\bibitem{shylesh2017study}
S.~Shylesh, ``A study of software development life cycle process models,'' in \emph{National Conference on Reinventing Opportunities in Management, IT, and Social Sciences}, 2017, pp. 534--541.

\bibitem{SODump}
S.~E. Community. (2023) Stack overflow data dump. [Online]. Available: \url{https://archive.org/details/stackexchange}

\bibitem{al2021empirical}
M.~A. Al~Alamin, S.~Malakar, G.~Uddin, S.~Afroz, T.~B. Haider, and A.~Iqbal, ``An empirical study of developer discussions on low-code software development challenges,'' in \emph{2021 IEEE/ACM 18th International Conference on Mining Software Repositories (MSR)}.\hskip 1em plus 0.5em minus 0.4em\relax IEEE, 2021, pp. 46--57.

\bibitem{g2}
G.~Team. (2023) G2.com. [Online]. Available: \url{https://www.g2.com/}

\bibitem{hadi2023survey}
M.~U. Hadi, R.~Qureshi, A.~Shah, M.~Irfan, A.~Zafar, M.~Shaikh, N.~Akhtar, J.~Wu, and S.~Mirjalili, ``A survey on large language models: Applications, challenges, limitations, and practical usage,'' \emph{TechRxiv}, 2023.

\bibitem{teubner2023welcome}
T.~Teubner, C.~M. Flath, C.~Weinhardt, W.~van~der Aalst, and O.~Hinz, ``Welcome to the era of chatgpt et al. the prospects of large language models,'' \emph{Business \& Information Systems Engineering}, vol.~65, no.~2, pp. 95--101, 2023.

\bibitem{orru2023human}
G.~Orr{\`u}, A.~Piarulli, C.~Conversano, and A.~Gemignani, ``Human-like problem-solving abilities in large language models using chatgpt,'' \emph{Frontiers in Artificial Intelligence}, vol.~6, p. 1199350, 2023.

\bibitem{park2023generative}
J.~S. Park, J.~C. O'Brien, C.~J. Cai, M.~R. Morris, P.~Liang, and M.~S. Bernstein, ``Generative agents: Interactive simulacra of human behavior,'' \emph{arXiv preprint arXiv:2304.03442}, 2023.

\bibitem{xu2023expertprompting}
B.~Xu, A.~Yang, J.~Lin, Q.~Wang, C.~Zhou, Y.~Zhang, and Z.~Mao, ``Expertprompting: Instructing large language models to be distinguished experts,'' \emph{arXiv preprint arXiv:2305.14688}, 2023.

\bibitem{wei2022chain}
J.~Wei, X.~Wang, D.~Schuurmans, M.~Bosma, F.~Xia, E.~Chi, Q.~V. Le, D.~Zhou \emph{et~al.}, ``Chain-of-thought prompting elicits reasoning in large language models,'' \emph{Advances in Neural Information Processing Systems}, vol.~35, pp. 24\,824--24\,837, 2022.

\bibitem{maxqda}
M.~Team. (2023) Maxqda. [Online]. Available: \url{https://www.maxqda.com/}

\bibitem{jetbrainsReport}
J.~Community. (2023) The state of developer ecosystem 2022. [Online]. Available: \url{https://www.jetbrains.com/lp/devecosystem-2022/}

\bibitem{quickbase}
 
Q.~Team. (2023) Quickbase. [Online]. Available: \url{https://www.quickbase.com/}
 

\bibitem{appypie}
 
A.~Team. (2023) Appypie. [Online]. Available: \url{https://www.appypie.com/}
 

\bibitem{appsheet}
 
------. (2023) Appsheet. [Online]. Available: \url{https://www.appsheet.com/}
 

\bibitem{softr}
 
S.~Team. (2023) Softr. [Online]. Available: \url{https://www.softr.io/}
 

\bibitem{appian}
 
A.~Team. (2023) Appian. [Online]. Available: \url{https://appian.com/}
 

\bibitem{servicenow}
 
S.~Team. (2023) Servicenow. [Online]. Available: \url{https://www.servicenow.com/}
 

\bibitem{airtable}
 
A.~Team. (2023) Airtable. [Online]. Available: \url{https://www.airtable.com/}
 

\bibitem{mendix}
 
M.~Team. (2023) Mendix. [Online]. Available: \url{https://www.mendix.com/}
 

\bibitem{pega}
 
P.~Team. (2023) Pega. [Online]. Available: \url{https://www.pega.com/}
 

\bibitem{nintex}
 
N.~Team. (2023) Nintex. [Online]. Available: \url{https://www.nintex.com/}
 

\bibitem{rao2023ai}
N.~Rao, J.~Tsay, K.~Kate, V.~J. Hellendoorn, and M.~Hirzel, ``Ai for low-code for ai,'' \emph{arXiv preprint arXiv:2305.20015}, 2023.

\bibitem{yen2023coladder}
R.~Yen, J.~Zhu, S.~Suh, H.~Xia, and J.~Zhao, ``Coladder: Supporting programmers with hierarchical code generation in multi-level abstraction,'' \emph{arXiv preprint arXiv:2310.08699}, 2023.

\bibitem{cai2023low}
Y.~Cai, S.~Mao, W.~Wu, Z.~Wang, Y.~Liang, T.~Ge, C.~Wu, W.~You, T.~Song, Y.~Xia \emph{et~al.}, ``Low-code llm: Visual programming over llms,'' \emph{arXiv preprint arXiv:2304.08103}, 2023.

\bibitem{autoGPT}
 
T.~et~al. (2023) Autogpt. [Online]. Available: \url{https://github.com/Significant-Gravitas/AutoGPT}
 

\bibitem{di2020democratizing}
C.~Di~Sipio, D.~Di~Ruscio, and P.~T. Nguyen, ``Democratizing the development of recommender systems by means of low-code platforms,'' in \emph{Proceedings of the 23rd ACM/IEEE international conference on model driven engineering languages and systems: companion proceedings}, 2020, pp. 1--9.

\bibitem{ihirwe2020low}
F.~Ihirwe, D.~Di~Ruscio, S.~Mazzini, P.~Pierini, and A.~Pierantonio, ``Low-code engineering for internet of things: a state of research,'' in \emph{Proceedings of the 23rd ACM/IEEE international conference on model driven engineering languages and systems: companion proceedings}, 2020, pp. 1--8.

\bibitem{oteyo2021building}
I.~N. Oteyo, A.~L.~S. Pupo, J.~Zaman, S.~Kimani, W.~De~Meuter, and E.~G. Boix, ``Building smart agriculture applications using low-code tools: the case for discopar,'' in \emph{2021 IEEE AFRICON}.\hskip 1em plus 0.5em minus 0.4em\relax IEEE, 2021, pp. 1--6.

\bibitem{martins2020overview}
R.~Martins, F.~Caldeira, F.~Sa, M.~Abbasi, and P.~Martins, ``An overview on how to develop a low-code application using outsystems,'' in \emph{2020 International Conference on Smart Technologies in Computing, Electrical and Electronics (ICSTCEE)}.\hskip 1em plus 0.5em minus 0.4em\relax IEEE, 2020, pp. 395--401.

\bibitem{zhuang2022easyfl}
W.~Zhuang, X.~Gan, Y.~Wen, and S.~Zhang, ``Easyfl: A low-code federated learning platform for dummies,'' \emph{IEEE Internet of Things Journal}, vol.~9, no.~15, pp. 13\,740--13\,754, 2022.

\bibitem{chen2022devicetalk}
W.-E. Chen, Y.-B. Lin, T.-H. Yen, S.-R. Peng, and Y.-W. Lin, ``Devicetalk: A no-code low-code iot device code generation,'' \emph{Sensors}, vol.~22, no.~13, p. 4942, 2022.

\bibitem{gomes2022low}
P.~M. Gomes and M.~A. Brito, ``Low-code development platforms: a descriptive study,'' in \emph{2022 17th Iberian Conference on Information Systems and Technologies (CISTI)}.\hskip 1em plus 0.5em minus 0.4em\relax IEEE, 2022, pp. 1--4.

\bibitem{rafi2022devops}
S.~Rafi, M.~A. Akbar, M.~S{\'a}nchez-Gord{\'o}n, and R.~Colomo-Palacios, ``Devops practitioners’ perceptions of the low-code trend,'' in \emph{Proceedings of the 16th ACM/IEEE International Symposium on Empirical Software Engineering and Measurement}, 2022, pp. 301--306.

\bibitem{luo2021characteristics}
Y.~Luo, P.~Liang, C.~Wang, M.~Shahin, and J.~Zhan, ``Characteristics and challenges of low-code development: the practitioners' perspective,'' in \emph{Proceedings of the 15th ACM/IEEE international symposium on empirical software engineering and measurement (ESEM)}, 2021, pp. 1--11.

\bibitem{sufi2023algorithms}
F.~Sufi, ``Algorithms in low-code-no-code for research applications: a practical review,'' \emph{Algorithms}, vol.~16, no.~2, p. 108, 2023.

\bibitem{rokis2022challenges}
K.~Rokis and M.~Kirikova, ``Challenges of low-code/no-code software development: A literature review,'' in \emph{International Conference on Business Informatics Research}.\hskip 1em plus 0.5em minus 0.4em\relax Springer, 2022, pp. 3--17.

\bibitem{chen2021evaluating}
M.~Chen, J.~Tworek, H.~Jun, Q.~Yuan, H.~P. d.~O. Pinto, J.~Kaplan, H.~Edwards, Y.~Burda, N.~Joseph, G.~Brockman \emph{et~al.}, ``Evaluating large language models trained on code,'' \emph{arXiv preprint arXiv:2107.03374}, 2021.

\bibitem{ross2023programmer}
S.~I. Ross, F.~Martinez, S.~Houde, M.~Muller, and J.~D. Weisz, ``The programmer’s assistant: Conversational interaction with a large language model for software development,'' in \emph{Proceedings of the 28th International Conference on Intelligent User Interfaces}, 2023, pp. 491--514.

\bibitem{liu2023wants}
M.~X. Liu, A.~Sarkar, C.~Negreanu, B.~Zorn, J.~Williams, N.~Toronto, and A.~D. Gordon, ``“what it wants me to say”: Bridging the abstraction gap between end-user programmers and code-generating large language models,'' in \emph{Proceedings of the 2023 CHI Conference on Human Factors in Computing Systems}, 2023, pp. 1--31.

\bibitem{kazemitabaar2023studying}
M.~Kazemitabaar, J.~Chow, C.~K.~T. Ma, B.~J. Ericson, D.~Weintrop, and T.~Grossman, ``Studying the effect of ai code generators on supporting novice learners in introductory programming,'' in \emph{Proceedings of the 2023 CHI Conference on Human Factors in Computing Systems}, 2023, pp. 1--23.

\bibitem{peng2023impact}
S.~Peng, E.~Kalliamvakou, P.~Cihon, and M.~Demirer, ``The impact of ai on developer productivity: Evidence from github copilot,'' \emph{arXiv preprint arXiv:2302.06590}, 2023.

\bibitem{nguyen2022empirical}
N.~Nguyen and S.~Nadi, ``An empirical evaluation of github copilot's code suggestions,'' in \emph{Proceedings of the 19th International Conference on Mining Software Repositories}, 2022, pp. 1--5.

\bibitem{vaithilingam2022expectation}
P.~Vaithilingam, T.~Zhang, and E.~L. Glassman, ``Expectation vs. experience: Evaluating the usability of code generation tools powered by large language models,'' in \emph{Chi conference on human factors in computing systems extended abstracts}, 2022, pp. 1--7.

\bibitem{liang2023large}
J.~T. Liang, C.~Yang, and B.~A. Myers, ``A large-scale survey on the usability of ai programming assistants: Successes and challenges,'' in \emph{2024 IEEE/ACM 46th International Conference on Software Engineering (ICSE)}.\hskip 1em plus 0.5em minus 0.4em\relax IEEE Computer Society, 2023, pp. 605--617.

\bibitem{sridhara2023chatgpt}
G.~Sridhara, S.~Mazumdar \emph{et~al.}, ``Chatgpt: A study on its utility for ubiquitous software engineering tasks,'' \emph{arXiv preprint arXiv:2305.16837}, 2023.

\bibitem{suri2023software}
S.~Suri, S.~N. Das, K.~Singi, K.~Dey, V.~S. Sharma, and V.~Kaulgud, ``Software engineering using autonomous agents: Are we there yet?'' in \emph{2023 38th IEEE/ACM International Conference on Automated Software Engineering (ASE)}.\hskip 1em plus 0.5em minus 0.4em\relax IEEE, 2023, pp. 1855--1857.





























































\end{thebibliography}



\end{document}